\documentclass[11pt,twoside,nohyper]{pallis}
\usepackage{rotating}
\usepackage{epsfig}
\usepackage{latexsym}
\usepackage{euscript}
\usepackage{amssymb}
\usepackage{fancyh}
\usepackage{times}
\usepackage{caption}
\usepackage{slashed,cite}
\usepackage{upgreek}

\usepackage{amsmath}
\usepackage{amsfonts}
\usepackage{amsbsy}
\usepackage{amscd}
\usepackage{bbm}

\newcommand{\ftn}{\footnotesize}

\newcommand{\GeV}{{\mbox{\rm GeV}}}
\newcommand{\ZeV}{{\mbox{\rm ZeV}}}
\newcommand{\EeV}{{\mbox{\rm EeV}}}
\newcommand{\PeV}{{\mbox{\rm PeV}}}
\newcommand{\YeV}{{\mbox{\rm YeV}}}

\newcommand{\bal}{\begin{align}}
\newcommand{\eal}{\end{align}}
\newcommand{\beqs}{\begin{subequations}}
\newcommand{\eeqs}{\end{subequations}}
\newcommand{\eec}{\end{center}}
\newcommand{\bec}{\begin{center}}

\newcommand{\ecs}{\end{cases}}
\newcommand{\bcs}{\begin{cases}}

\newcommand{\eem}{\end{matrix}}
\newcommand{\bem}{\begin{matrix}}
\newcommand{\eeq}{\end{equation}}
\newcommand{\beq}{\begin{equation}}
\newcommand{\ba}{\begin{array}}
\newcommand{\ea}{\end{array}}
\newcommand{\bea}{\begin{eqnarray}}
\newcommand{\eea}{\end{eqnarray}}
\newcommand{\baq}{\begin{eqnarray}}
\newcommand{\eaq}{\end{eqnarray}}

\newcommand{\sFref}[2]{Fig.~\ref{#1}-{\small\sf ({#2})}}
\newcommand{\sEref}[2]{Eq.~(\ref{#1}{{\small\sf  #2}})}

\newcommand{\Eref}[1]{Eq.~(\ref{#1})}

\newcommand{\Sref}[1]{Sec.~\ref{#1}}
\newcommand{\Srefs}[1]{Secs.~\ref{#1}}
\newcommand{\Fref}[1]{Fig.~\ref{#1}}
\newcommand{\Tref}[1]{Table~\ref{#1}}
\newcommand{\cref}[1]{Ref.~\cite{#1}}
\newcommand{\crefs}[1]{Refs.~\cite{#1}}
\newcommand{\etal}{{\it et al.\/}}
\renewcommand{\email}[1]{{{\sl e-mail address:}~{\tt #1}}}

\newcommand\eqs[2]{Eqs.~(\ref{#1}) and (\ref{#2})}
\newcommand\eqss[3]{Eqs.~(\ref{#1}), (\ref{#2}) and (\ref{#3})}

\newcommand{\ffm}{\ensuremath{F_{\rm T}}}
\newcommand{\ffe}{\ensuremath{F_{\rm sh}}}
\newcommand{\ffd}{\ensuremath{F_{\rm d}}}
\newcommand{\ffw}{\ensuremath{F_{\rm W}}}
\newcommand{\fm}{\ensuremath{f_{\rm T}}}
\newcommand{\fw}{\ensuremath{f_{\rm W}}}
\newcommand{\fe}{\ensuremath{f_{\rm sh}}}
\newcommand{\fp}{\ensuremath{f_{\rm T}}}
\newcommand{\fps}{\ensuremath{f_{\rm T\star}}}
\newcommand{\fd}{\ensuremath{f_{\rm d}}}
\newcommand{\ks}{\ensuremath{k_\star}}

\newcommand{\nus}{\ensuremath{N_S}}
\newcommand{\nd}{\ensuremath{n_{\rm d}}}
\newcommand{\nm}{\ensuremath{N}}

\renewcommand{\Im}{{\mbox{\sf\small Im}}}
\renewcommand{\Re}{{\mbox{\sf\small Re}}}

\newcommand\vev[1]{\langle {#1} \rangle}
\newcommand\vevi[1]{\langle {#1} \rangle_{\rm I}}
\def\lf{\left(}
\def\rg{\right)}

\def\llgm{\left\lgroup}
\def\rrgm{\right\rgroup}

\newcommand{\Vhi}{\ensuremath{V_{\rm I}}}
\newcommand{\vf}{\ensuremath{V_{\rm F}}}

\newcommand{\Vci}{\ensuremath{V_{\rm I}}}
\newcommand{\dV}{\ensuremath{\Delta V_{\rm I}}}
\newcommand{\Hhi}{\ensuremath{H_{\rm I}}}
\newcommand{\as}{\ensuremath{\alpha_{\rm s}}}
\newcommand{\wrh}{\ensuremath{w_{\rm rh}}}

\newcommand{\Dex}{\ensuremath{\Delta_{\star}}}

\newcommand{\rcc}{\ensuremath{\mathcal R}}

\newcommand{\Ve}{\ensuremath{V}}

\newcommand{\plk}{{\slshape\small Planck}}

\def\bbet{{\bar\beta}}
\def\al{{\alpha}}
\def\bt{{\beta}}

\def\n{\bar{n}}

\newcommand{\Trh}{\ensuremath{T_{\rm rh}}}

\newcommand{\ld}{\ensuremath{\lambda}}
\newcommand{\Ld}{\ensuremath{\Lambda}}

\newcommand{\kp}{\ensuremath{\kappa}}
\newcommand{\kpp}{\ensuremath{\kappa_+}}
\newcommand{\kpm}{\ensuremath{\kappa_-}}
\newcommand{\se}{\ensuremath{\widehat\phi}}
\newcommand{\sex}{\ensuremath{\widehat{\phi}_*}}
\newcommand{\sgx}{\ensuremath{\phi_\star}}
\newcommand{\Ldx}{\ensuremath{\Lambda_\star}}
\newcommand{\Hx}{\ensuremath{H_{\rm I\star}}}
\newcommand{\sgm}{\ensuremath{\phi_{\rm mx}}}
\newcommand{\sgn}{\ensuremath{\phi_{\rm mn}}}

\newcommand{\sef}{\ensuremath{\widehat{\phi}_{\rm f}}}

\newcommand{\eph}{\ensuremath{\epsilon}}
\newcommand{\ith}{\ensuremath{\eta}}
\newcommand{\mP}{\ensuremath{m_{\rm P}}}

\newcommand{\sg}{\ensuremath{\phi}}

\renewcommand{\sigma}{\ensuremath{\phi}}

\newcommand{\sgf}{\ensuremath{\phi_{\rm f}}}

\newcommand{\what}{\ensuremath{\widehat}}
\newcommand{\Khi}{\ensuremath{K}}

\newcommand{\diag}{\mbox{\sf\small diag}}

\newcommand{\Ggut}{\ensuremath{G_{\rm GUT}}}
\newcommand{\Gsm}{\ensuremath{G_{\rm SM}}}

\newcommand{\mbl}{\ensuremath{M_{X}}}
\newcommand{\mgut}{\ensuremath{M_{\rm GUT}}}
\newcommand{\bl}{\ensuremath{U(1)_{X}}}
\newcommand{\ldu}{\ensuremath{\uplambda}}
\newcommand{\ns}{\ensuremath{n_{\rm s}}}
\newcommand{\As}{\ensuremath{A_{\rm s}}}
\newcommand{\Ns}{\ensuremath{N_{\star}}}

\newcommand\mtt[4]{\mbox{
$\llgm\bem #1 &#2 \cr #3& #4\eem\rrgm$}}

\newcommand\stl[2]{\mbox{$\llgm\bem #1\cr #2\eem\rrgm$}}

\newcommand{\phc}{\ensuremath{\Phi}}
\newcommand{\phcb}{\ensuremath{\bar\Phi}}
\newcommand{\psc}{\ensuremath{\Psi}}
\newcommand{\pscb}{\ensuremath{\bar\Psi}}
\newcommand{\php}{\ensuremath{\Phi_+}}
\newcommand{\phm}{\ensuremath{\Phi_-}}

\newcommand\mtta[4]{\mbox{
$\llgm\bem #1 &#2 \cr #3& #4\eem\rrgm$}}

\newcommand{\dn}{\ensuremath{\delta n}}

\newcommand{\rs}{\ensuremath{\delta_{21}}}

\renewcommand{\Dex}{\ensuremath{\Delta_{\rm \star}}}
\newcommand{\Lex}{\ensuremath{\Lambda_{\rm \star}}}

\newcommand{\kbaad}{\ensuremath{{\widetilde  K_{2(11)^2\rm d}}}}
\newcommand{\kbbad}{\ensuremath{{\widetilde  K_{221\rm d}}}}
\newcommand{\kb}{\ensuremath{K_{2}}}

\newcommand{\tkaa}{\ensuremath{{\widetilde K_{(11)^2}}}}

\newcommand{\tkbah}{\ensuremath{{\widetilde K_{21}}}}

\newcommand{\tkbaad}{\ensuremath{{\widetilde K_{2(11)^2\rm d}}}}
\newcommand{\tkbbad}{\ensuremath{{\widetilde K_{221\rm d}}}}
\newcommand{\tkbad}{\ensuremath{{\widetilde K_{211\rm d}}}}

\newcommand{\kaa}{\ensuremath{{K_{(11)^2}}}}
\newcommand{\tka}{\ensuremath{{\widetilde K_{11}}}}

\newcommand{\tkm}{\ensuremath{\widetilde K_{\rm T}}}

\newcommand{\tkbmd}{\ensuremath{\widetilde K_{2\rm Td}}}

\newcommand{\km}{\ensuremath{K_{\rm T}}}

\newcommand{\kd}{\ensuremath{K_{\rm d}}}
\newcommand{\krms}{\ensuremath{K_{\rm s}}}
\newcommand{\ksh}{\ensuremath{K_{\rm sh}}}

\newcommand{\mnfaa}{\ensuremath{\mathcal{M}_{(11)^2}}}

\newcommand{\mnfbba}{\ensuremath{\mathcal{M}_{221}}}
\newcommand{\mnfbaa}{\ensuremath{\mathcal{M}_{2(11)^2}}}
\newcommand{\mnfba}{\ensuremath{\mathcal{M}_{211}}}
\newcommand{\mnfbah}{\ensuremath{\mathcal{M}_{21}}}

\def\Ka{K\"{a}hler~}
\def\Km{K\"{a}hler manifold}
\def\Kme{K\"{a}hler metric}
\def\Kap{K\"{a}hler potential}
\def\sub{subplanckian}

\newcommand{\dphi}{\ensuremath{\what{\delta\phi}}}
\newcommand{\dph}{\ensuremath{\delta\phi}}
\newcommand{\msn}{\ensuremath{\what m_{\rm I}}}

\renewenvironment{subequations}{%
\refstepcounter{equation}%
\setcounter{parentequation}{\value{equation}}%
  \setcounter{equation}{0}
  \def\theequation{\thesection.\theparentequation{\sf\ftn \alph{equation}}}%
  \ignorespaces
}{%
  \setcounter{equation}{\value{parentequation}}%
  \ignorespacesafterend
}

\title{\LARGE\boldmath \bfseries\scshape Starobinsky Inflation
with T-Model K\"ahler Geometries}

\author{\Large \bfseries\scshape C. Pallis\\
School of Technology, \\ Aristotle
University of Thessaloniki, \\ GR-541 24 Thessaloniki, GREECE \\
\vspace{3pt}
\email{kpallis@auth.gr}}

\abstract{We present novel implementations of Starobisky-like
inflation within Supergravity adopting \Kap s for the inflaton
which parameterize hyperbolic geometries known from the T-model
inflation. The associated superpotentials are consistent with an
$R$ and a global or gauge $U(1)_X$ symmetries. The inflaton is
represented by a gauge-singlet or non-singlet superfield and is
accompanied by a gauge-singlet superfield successfully stabilized
thanks to its compact contribution into the total K\"ahler
potential. Keeping the \Km\ intact, a conveniently violated shift
symmetry is introduced which allows for a slight variation of the
predictions of Starobinsky inflation: The (scalar) spectral index
exhibits an upper bound which lies close to its central
observational value whereas the constant scalar curvature of the
inflaton-sector K\"ahler manifold increases with the
tensor-to-scalar ratio.

\\ \\ {\ftn\sffamily {\scshape Keywords}:  Cosmology, Inflation, Supersymmetric Models} \\
{\ftn\sffamily {\scshape PACS codes}:  98.80.Cq, 12.60.Jv,
95.30.Cq, 95.30.Sf}
\\\\ {\sl\bfseries Published in} {\sl Universe} {\bf 11}, no.~3,
75 (2025) \\ {\sf\small Special Issue} \\ {\sl Particle Physics
and Cosmology: A Themed Issue in Honor of Professor Dimitri
Nanopoulos}}

\begin{document}

\setcounter{page}{1} \pagestyle{fancyplain}

\addtolength{\headheight}{.5cm}

\rhead[\fancyplain{}{ \bf \thepage}]{\fancyplain{}{\sc SI with
T-Model K\"ahler Geometries}} \lhead[\fancyplain{}{\sc
\leftmark}]{\fancyplain{}{\bf \thepage}} \cfoot{}

\section{Introduction}\label{intro}

\emph{Starobinsky inflation} ({\sf\small SI}) \cite{str} stands
out among the remaining viable inflationary models (for reviews
see \cref{rev, rev1, rev2,plin}) thanks to its simplicity,
elegance and observational success. Despite its original
realization by the (arbitrary) addition of the $\rcc^2$ term --
where $\rcc$ is the Ricci scalar -- to the standard Einstein
action, this inflationary model can be also driven by a scalar
field with a suitable potential. Indeed, it is well known that
gravity theories based on higher derivative terms of the type
$\rcc^m$ with $m>1$ are equivalent to standard gravity theories
with one additional scalar degree of freedom \cite{nick}. From the
proposed particle-physics incarnations of this inflationary model
(and its variant called $\alpha$-SI) -- for reviews see
\cref{nsreview, class, prl, su11a} -- prominent position occupies
its embedding within \emph{Supergravity} ({\sf\small SUGRA}) which
is the natural extension of \emph{Supersymmetry} ({\sf\small
SUSY}) to planckian mass scales \cite{gref}. Despite the fact that
SUSY is not yet discovered at LHC \cite{lhc}, its presence --
probably at higher energies -- is a natural and mostly inevitable
consequence of superstring theory \cite{susy22}. Even without
direct experimental signatures, SUSY has constructive impact on
the stabilization of the electroweak vacuum and on several
problems of modern Particle Cosmology such as inflation,
baryogenesis and dark matter.

Trying to classify the most popular SUGRA realizations of SI we
can single out indicatively the following categories -- for
alternatives see, e.g., \cref{nick,ketov,bas,buch,stra,strb}:

\begin{itemize}

\item Wess-Zumino models with a matter-like inflaton
\cite{eno5,eno7,gut1,gut2,gut5f,gut10}. Polynomial
superpotentials, $W$, -- of the Wess-Zumino form \cite{gref} --
are adopted in this class of models and the \Kap s $K$
parameterize specific \Km s of the form $SU(N_{\rm m},1)/SU(N_{\rm
m})\times U(1)$, inspired by the no-scale models
\cite{noscale,lahanas} of SUSY breaking. The stabilization of the
inflaton-accompanying modulus at a Planck-scale value \cite{eno7}
is achieved by a deformation of the internal geometry.


\item Ceccoti-like \cite{cecoti} models with a modulus-like
inflaton \cite{linder2, R2r, nIG, nIGpl, su11, farakos,eno7,rena,
tamv, ighi, blst}. Similar $K$'s are used here whereas $W$ is
linear \cite{rube} with respect to the matter-like
inflaton-accompanying field which may be stabilized at the origin
via several mechanisms \cite{nick, lee, su11,
su11a,nil,nila,nilb,ano}. In a subclass of these models \cite{R2r,
nIG, nIGpl, rena, ighi, blst}, the conjecture of induced gravity
\cite{zee,gian} is incorporated leading to a dynamical generation
of the reduced Planck scale, $\mP$ through the \emph{vacuum
expectation value} ({\sf\small v.e.v}) of the inflaton at the end
of its evolution.

\item Models with a strong linear non-minimal coupling to gravity
\cite{quad1,quad,unh} -- or a strong linear contribution into this
coupling \cite{unh1} -- which remain unitarity safe \cite{riotto}
up to the Planck scale.

\item Models which exhibit a pole
\cite{terada,pole1,pole2,polec,polena} of order \emph{one} in the
kinetic term of the inflaton \cite{tkref,tkrefa,epole}. As in
every SI model, the inflationary potential develops one shoulder
for large $\se$ values, where $\se$ is the canonically normalized
inflaton which can be expressed in terms of the original field
$\sg$ as \cite{epole}
\beq \sg=1-e^{-\sqrt{2/N}\se}~~\big(\mbox{: E-Model
Normalization}\big).\label{emd} \eeq
The presence of the real positive variable $N$ -- aligned with the
conventions of \cref{epole} -- leads to a generalized version of
SI called $\alpha$-SI \cite{eno7} or \emph{E-Model} inflation.
This model can be contrasted with the \emph{T-Model} inflation
\cite{tmodel,alinde} which arises thanks to a pole of order
\emph{two} in the inflaton kinetic term and features a potential
with two symmetric plateaus away from the origin. Namely, the
$\sg-\se$ relation assumes the form
\beq \sg= \tanh{\lf\se/\sqrt{2N}\rg}~~\big(\mbox{: T-Model
Normalization}\big).\label{tmd}\eeq
Independently of their particularities, both models share
\cite{class} common predictions for the inflationary observables
and for this reason they are called collectively
$\alpha$-attractors \cite{plin, linde21,ellis21,attr21}.

\end{itemize}

In our present investigation, stimulated by \cref{tmodel} -- see
also \cref{dimbook} --, we propose another embedding of SI within
${\cal N}=1$ standard Poincar\'e  SUGRA  which is exclusively
based on the presence of a pole of order \emph{two} in the
inflaton kinetic term, i.e., it is based on the $\sg-\se$ relation
of \Eref{tmd}. Namely, the scalar potential, expressed in terms of
the initial (non-canonical) inflaton, $\sg$, is written as
\beq V_{\rm I}=\ld^2(\sg^{n/2}-M^2)^2/(1+\sg)^{\nd}
\>\>\>\mbox{with}\>\>\>M\ll\mP=1.\label{vhi}\eeq
For $n=\nd=2$, $\Vhi$ has been motivated in \cref{linder2,tmodel}
via a breaking of conformal symmetry in a non-SUSY framework. It
also appears in unified models of no-scale $\alpha$-SI with SUSY
breaking \cite{unified}. Here we extend the analysis in the
context of SUGRA providing a method which allows the generation of
$\Vhi$ from conveniently selected $W$ and $K$. Our particle
content includes, besides the inflaton superfield(s), an extra
gauge-singlet superfield which assist in the stabilization of the
SUGRA potential during SI \cite{rube}. Moreover, we employ
monomial $W$'s consistent with an $R$ and a global or gauge $\bl$
symmetry. On the other hand, the $K$'s respect the $R$ and the
gauge symmetry and are \emph{holomorphically} equivalent to those
yielding \Eref{tmd} \cite{tkref, sor, sor1, sor2}. In other words,
$K$ has the well-known form employed in T-model inflation
\cite{tkref, sor} up to a number of extra holomorphic and
anti-holomorphic terms which do not influence the resulting \Kme.
The one pair of these terms endows $K$ with a shift symmetry which
facilitates the performance of inflation -- cf.
\cref{tkref,sor,sor1,sor2}. The second pair provides a breaking of
the aforementioned shift symmetry which is already violated mildly
in $W$. Both violations are natural in the 't Hooft sense
\cite{symm} for low enough values of the exponent $\nd$ in $K$ and
the coupling constant $\ld$ entered in $W$. The particular
importance of an enhanced shift symmetry in taming the so-called
$\eta$-problem of inflation in SUGRA is already recognized for
gauge singlets, e.g., in
\cref{kawasaki,crk,nmkine,nanoli,brax,antu} and non-singlets,
e.g., in \cref{shiftH, jhep, nmkinh, nmkinha, unh, ighi}.


Compared to the aforementioned implementations of SI within SUGRA,
the present version assures a totally symmetric \Km\ together with
a simple $W$. Let us recall that polynomial $W$'s with $R$
symmetry are reconciled with a $K$ which does not parameterize
specific \Km\ in \cref{epole} whereas totally symmetric $K$'s
usually require complicate $W$'s -- see e.g. \cref{eno7, tkref,
tkrefa, linde21,tamv}. On the contrary, the employed here $W$'s
are very common in particle physics and so the inflaton can be
easily identified with a field already present in the theory, e.g,
the right-handed sneutrino \cite{kawasaki, sneut, sneut1, sneut2}
or a superheavy Higgs superfield responsible for the spontaneous
breaking of a gauge symmetry \cite{ighi,blst}. Also, it is
expected that this scheme assures naturally a low enough reheating
temperature, potentially consistent with the gravitino constraint
and non-thermal leptogenesis \cite{R2r,ighi,blst,phenoAt,uninano}
if connected with a version of the \emph{Minimal SUSY Standard
Model} ({\sf\small MSSM}). Furthermore, our proposal does not
require tuning of parameters -- besides a mild one in the initial
conditions --, it is not based on any conjecture such as that of
induced gravity \cite{R2r,nIG,nIGpl,ighi, blst, rena,su11a,su11}
and provides a relative flexibility as regards the observables. At
last, it offers us the opportunity to exemplify the \Kap\
engineering which allows to obtain the various desired factors of
$\Vhi$ in \Eref{vhi} together with a desirable \Kme.

We below describe how we can formulate SI in the context of SUGRA
in \Sref{sugra} and we specify two versions of SI: one employing a
gauge-singlet inflaton in \Sref{ci} called for short \emph{Chaotic
Starobinsky Inflation} ({\sf\small CSI}) and one with a
gauge-non-singlet inflaton called \emph{Higgs Starobinsky
Inflation} ({\sf\small HSI}) since it causes the breaking of a
gauge symmetry -- see \Sref{hi}. Our conclusions are summarized in
\Sref{con}. In Appendix \ref{app} we demonstrate that our proposed
$K$'s enjoy an enhanced shift symmetry. Throughout the text, the
subscript $,\chi$ denotes derivation \emph{with respect to}
({\sf\small w.r.t}) the field $\chi$ and charge conjugation is
denoted by a star ($^*$). Unless otherwise stated, we use units
where the reduced Planck scale $\mP = 2.4\cdot 10^{18}~\GeV$ is
set equal to unity.

\section{SUGRA Framework}\label{sugra}

We start our investigation presenting the basic formulation of a
scalar theory within SUGRA in \Sref{sugra1} and then -- in
\Sref{sugra2} -- we outline our strategy in constructing our
models of SI.

\subsection{General Set-up} \label{sugra1}

The part of the (Einstein-frame) action within SUGRA which
describes the (complex) scalar fields $z^\al$ coupled minimally to
Einstein gravity can be written as \cite{gref}
\beqs\beq \label{action1} {\cal A} = \int d^4 x
\sqrt{-\mathfrak{g}} \left(-\frac{1}{2}{\rcc} +K_{\al\bbet} D_\mu
z^\al D^\mu z^{*\bbet}-V_{\rm SUGRA}\right)\,, \eeq
where $\rcc$ is the space-time Ricci scalar curvature,
$\mathfrak{g}$ is the determinant of the
Friedmann-Robertson-Walker metric, $g_{\mu\nu}$, with signature
$(+,-,-,-)$.  Also, summation is taken over the scalar fields
$z^\al$ which are denoted by the same symbol of the corresponding
superfield. The kinetic mixing of $z^\al$ is controlled by the
K\"ahler potential $K$ and the relevant metric defined as
\beq \label{kddef} K_{\al\bbet}={\Khi_{,z^\al
z^{*\bbet}}}>0\>\>\>\mbox{with}\>\>\>K^{\bbet\al}K_{\al\bar
\gamma}=\delta^\bbet_{\bar \gamma}.\eeq
Also, the covariant derivatives for the $z^\al$'s are given by
\beq D_\mu z^\al=\partial_\mu z^\al+ig A^{\rm a}_\mu T^{\rm
a}_{\al\bt} z^\bt\eeq
with $A^{\rm a}_\mu$ being the vector gauge fields, $g$ the
(unified) gauge coupling constant and $T^{\rm a}$ with ${\rm
a}=1,...,\mbox{\sf\ftn dim}\Ggut$ the generators of a gauge group
$\Ggut$.

Finally ${\cal A}$ contains the SUGRA scalar potential, $V_{\rm
SUGRA}$, which is given in terms of $K$, and the superpotential,
$W$, by
\beq V_{\rm SUGRA}=\Ve_{\rm F}+ \Ve_{\rm D}\>\>\>\mbox{with}\>\>\>
\Ve_{\rm F}=e^{\Khi}\left(K^{\al\bbet}{\rm F}_\al {\rm
F}^*_\bbet-3{\vert W\vert^2}\right) \>\>\>\mbox{and}\>\>\>\Ve_{\rm
D}= g^2 \sum_{\rm a} {\rm D}_{\rm a} {\rm D}_{\rm a}/2,
\label{Vsugra} \eeq
where a trivial gauge kinetic function is adopted whereas the F-
and D-terms read
\beq \label{Kinv} {\rm F}_\al=W_{,z^\al}
+K_{,z^\al}W\>\>\>\mbox{and}\>\>\>{\rm D}_{\rm a}= z_\al\lf T_{\rm
a}\rg^\al_\bt K^\bt\>\>\>\mbox{with}\>\>\>
K^{\al}={\Khi_{,z^\al}}\,.\eeq\eeqs
As we emphasized in \Sref{intro}, SI in our work is attained by
deriving $\Vhi$ in \Eref{vhi} from $V_{\rm SUGRA}$ and not
modifying gravity which remains at the minimal level as shown from
the absence of higher order $\rcc$ terms in \Eref{action1}.
Therefore our next task is to select conveniently the functions
$K$ and $W$ so that \eqs{tmd}{vhi} are reproduced.

\subsection{Guidelines}\label{sugra2}

We embark on describing our procedure to obtain the desired $\Vhi$
in \Eref{vhi} from $V_{\rm F}$ in \Eref{Vsugra} and the desired
$\sg-\se$ relation in \Eref{tmd}. Although our presentation in
adapted to our present model, the strategy of our approach has a
wider applicability suitable for other cases too.

\subsubsection{Achieving D-Flatness.}\label{sugra2a}

Our final aim is the derivation of $\Vhi$ in \Eref{vhi} through
$V_{\rm F}$ in \Eref{Vsugra}. This decision requires the
establishment of D-flatness during SI, i.e., $\vevi{V_{\rm D}}=0$
-- where $\vevi{Q}$ symbolizes the value of the quantity $Q$
during inflation. Assuming that the gauge non-singlet superfields
are placed at zero during inflation, D-flatness may be attained in
the following two cases:

\begin{itemize}
\item If the inflaton is (the radial part of) a gauge-singlet
superfield $z^2:=\Phi$. In this case, $\Phi$ has obviously zero
contribution to $V_{\rm D}$.

\item If the inflaton is the radial part of a conjugate pair of
Higgs superfields, $z^2:=\Phi$ and $z^3:=\bar\Phi$, in the
fundamental representation of $\Ggut$. In a such case -- see
\Sref{hi2} below -- we obtain $\vevi{V_{\rm D}}=0$. The same
result can be obtained if $\Ggut$ is more structured than $\bl$
employing just one superfield $z^2$ in the adjoint representation
of $\Ggut$ and using as inflaton its neutral component -- see e.g.
\cref{jhepcs}.

\end{itemize}

\subsubsection{Selecting the Suitable $W$.}\label{sugra2b}

Despite the fact that $\Vhi$ in \Eref{vhi} includes only one
field, its derivation from $\vf$ is facilitated \cite{rube} if $W$
includes at least two fields from which the first $z^1:=S$ is a
gauge-singlet superfield, called stabilizer or goldstino. The
latter is due to the fact that for $S=0$ SUSY is broken since
$\vevi{F_S}\neq0$. The presence of $S$ serves the following
purposes:

\begin{itemize}

\item It assists in determining $W$. To achieve it, we require
that $S$ appears linearly in $W$ and so both are equally charged
under a global $R$ symmetry.

\item It can be stabilized at $\vevi{S}=0$ without invoking higher
order terms, if we select \cite{su11}
\beq \label{kb}
\kb=\nus\ln\lf1+|S|^2/\nus\rg~~\Rightarrow~~\vevi{K_2^{SS^*}}=1~~~\mbox{with}
~~~0<N_2<6.\eeq
The index 2 stems from the fact that $\krms$ parameterizes the
compact manifold $SU(2)/U(1)$ \cite{su11}. Note that for
$\vevi{S}=0$, $S$ is canonically normalized and so we do not care
about its kinetic normalization henceforth. For other
stabilization methods of $S$ see \cref{lee, nil, nila, nilb, nick,
ano}.

\item It assures the boundedness of $\Vhi$. Indeed, if we set
$\vevi{S}=0$, then $\vevi{W_{,z^\al}}=0$ for $\al\neq1$,
$\vevi{K_{,z^\al}W}=0$  and $-3\vert \vevi{W}\vert^2=0$.
Obviously, non-vanishing values of the last term may render
$\Ve_{\rm F}$ unbounded from below.

\item It generates for $\vevi{S}=0$ and for monomial $W$ the
numerator of $\Vhi$ in \Eref{vhi} via the only term of $V_{\rm F}$
which remains ``alive''. Indeed, we obtain
\beq \Vhi:=\vevi{V_{\rm F}}= \vevi{e^{K}K^{SS^*}|W_{,S}|^2}.
\label{Vhio}\eeq
Assuming that no mixing terms between $S$ and the inflaton exist
in $K$, we obtain $\vevi{K^{SS^*}}=\vevi{K_2^{SS^*}}=1$ and so the
numerator of $\Vhi$ in \Eref{vhi} emerges if $W$ has the form
\beq\label{ffw}
W=S\ffw(z^\al)~~~\mbox{with}~~~\vevi{\ffw}=:\fw=\ld\phi^{n/2}~~~\mbox{and}~~~\phi=\Re\phc,\eeq
given that the assumption $\vevi{\Im\phc}=0$ yields mostly stable
configuration -- here we focus on a gauge-singlet $\phc$. On the
other hand, the denominator of $\Vhi$ in \Eref{vhi} may be
generated via the exponential prefactor in \Eref{Vhio} through
logarithmic contributions to $K$ -- as we explain below.
\end{itemize}


\subsubsection{Selecting the Convenient \Ka\ Potential.} \label{sugra2c}

The form of $K$ has to accomplish the following two goals:

\begin{itemize}

\item It has to generate the desired $\sg-\se$ relation in
\Eref{tmd}. Therefore, we need to introduce a contribution into
$K$ including  $z^\al$ and $z^{*\al}$ in the same function. After
inspection -- see Appendix of \cref{ethybrid} -- we infer that a
pole of order two in the kinetic term of inflaton is achieved if
$K=\km$ where
\beq \km= -\nm
\ln\ffm(z^\al,z^{*\bbet})~~~\mbox{with}~~~\vevi{K_{\al\bbet}}=\nm/\fm^2
~~~\mbox{and}~~~\fm:=\vevi{\ffm}=1-\sg^2.\label{ffm}\eeq
Here $\nm>0$ and the subscript ``T'' indicates that this part of
the total $K$ is responsible for the T-model \Kme\ -- see
\Eref{kddef}. However, from \Eref{Vhio}, we remark that $K$
affects -- besides the kinetic mixing -- $\Vhi$ via the prefactor
$e^{\km}$. Therefore, $\ffm$ is generically expected to emerge
also in the denominator of $\Vhi$ making difficult the
establishment of an inflationary era. This problem can be
surpassed \cite{sor, epole} by two alternative strategies:

\begin{itemize}

\item Adjusting $W$ and constraining the prefactor of $K$ in
\Eref{ffm}, so that the pole is removed from $V_{\rm F}$ thanks to
cancellations \cite{sor,epole,eno5,polec}. This recipe introduces
some tuning, though, in the coefficients of $W$ and, for this
reason, we do not pursue this method here.

\item Replacing $\km$ with $\tkm$ so that the desired kinetic
terms in \Eref{action1} remain unaltered and, simultaneously
\cite{tkref, sor, epole}
\beqs\beq \vevi{e^{\tkm}}=1~\Leftrightarrow~
\vevi{\tkm}=0~~~\mbox{with}~~~\tkm=\km+\ksh.\label{ksh}\eeq
In other words, the symmetry of $\km$ is augmented by some shift
symmetry -- see Appendix~\ref{app} -- without disturbing
$K_{\al\bbet}$ in \Eref{kddef}. To accomplish this, $\ksh$
includes holomorphic and anti-holomorphic terms which yield
vanishing contribution to the mixed derivatives of $\tkm$. Taking
into account the form of $\km$ in \Eref{ffm}, we may select
formally
\beq
\ksh=({\nm}/{2})\ln\ffe+({\nm}/{2})\ln\ffe^*~~~\mbox{with}~~~\vevi{\ffe}=:\fe=\fm.\label{ffe}\eeq\eeqs
Note that the same construction is valid even in case of
polynomial $K$'s if we check the structure of the relevant $K$'s
in \cref{kawasaki, shiftH, crk, jhep, nmkinha, nmkinh, nmkine}.

\end{itemize}

\item It has to generate the denominator of $\Vhi$ in \Eref{vhi}.
To achieve this, we focus on the exponential prefactor of $\Vhi$
in \Eref{Vhio} and we demand
\beqs\beq \vevi{e^{\kd}}=(1+\sg)^{-\nd},\label{kd}\eeq
where $\kd$ has the following structure (similar to that of
$\ksh$)
\beq\kd=-(\nd/2)\ln\ffd-(\nd/2)\ln\ffd^*~~~\mbox{with}~~~\vevi{\ffd}=:\fd=1+\sg.\label{ffd}\eeq\eeqs
so that it does not disrupt $K_{\al\bbet}$ in \Eref{kddef}.

\end{itemize}

\paragraph{} To recapitulate this section, we summarize that the
adopted $W$ in this work has the form in \Eref{ffw} whereas the
total $K$ can be written as
\beq
\tkbmd=\kb+\tkm+\kd~~~\mbox{with}~~~\tkm=\km+\ksh.\label{tk}\eeq
Below, we specify the functional forms of the related functions
$\ffm, \ffe$ and $\ffd$ for the two classes of SI considered, CSI
and HSI.

\section{Gauge-Singlet Inflaton}\label{ci}

We focus first on the case of CSI and we present in \Sref{ci0} the
building blocks of the model. Then we verify that the adopted $K$
and $W$ produce the desired kinetic mixing in \Sref{ci1} and
inflationary potential in \Sref{ci2}.

\subsection{Set-up}\label{ci0}

According to the strategy described in \Sref{sugra2} the present
setting is realized in presence of two gauge-singlet superfields
$S$ and $\Phi$ which may be parameterized as
\beq \Phi=\:{\phi\,e^{i \theta}}\>\>\>\mbox{and}\>\>\>S=\:(s
+i\bar s)/\sqrt{2}.\label{para} \eeq
The inflationary trajectory can be defined by the constraints
\beq \label{inftr}
\vevi{S}=\vevi{\Phi-\Phi^*}=0,\>\>\mbox{or}\>\>\>\vevi{s}=\vevi{\bar
s}=\vevi{\theta}=0.\eeq

Following our plan in \Sref{sugra2b} we adopt a monomial $W$
consistent with \Eref{ffw} with the following structure
\beq \ffw=\ld\phc^{n/2} ~~~\mbox{and so}~~~ W= \ld S\phc^{n/2},
\label{wci} \eeq
where $\ld$ is a free parameter and $n$ takes even values which
preserve the holomorphicity of $W$.  The form of $W$ can be
uniquely determined if we impose an $R$ symmetry, under which $S$
and $\Phi$ have charges $1$ and $0$, and a global $\bl$ symmetry
with assigned charges $Q_X(S)=-1$ and $Q_X(\phc)=2/n$. The latter
is violated though in the proposed $K$ which assumes the form in
\Eref{tk} with the functions defined in \eqss{ffm}{ffe}{ffd}
identified as follows
\beq \ffm=1-|\phc|^2,~\ffe=1-\phc^2~~\mbox{and}~~\ffd=1+\Phi.
\label{ffmed}\eeq
Consequently, the total $K$ takes the form
\beq \tkbad=\kb+\tka+\kd, \label{tkbad}\eeq
where the individual contributions are specified as
\beq\tka=-\nm\ln\frac{1-|\phc|^2}{\sqrt{(1-\phc^2)(1-\phc^{*2})}}~~~\mbox{and}~~~
K_{\rm
d}=-\frac{\nd}{2}\ln(1+\Phi)-\frac{\nd}{2}\ln(1+\Phi^*).\label{tkad}\eeq
$\tkbad$ parameterizes \cite{tkref} the \Km\ \mnfba\ with
moduli-space scalar curvature ${\sf R}_{211}$ given respectively
by
\beq \label{mnfs} \mnfba=(SU(2)/U(1))_S\times \lf
SU(1,1)/U(1)\rg_{\phc}\>\>\>\mbox{and}\>\>\>{\sf
R}_{211}=2/\nus-{2}/{N}. \eeq
The indices in the product of \Eref{mnfs} indicate the moduli
which parameterize the corresponding manifolds. Needless to say,
$\kd$ and the denominator in \tka\ have no impact on $\mnfba$, as
explained in \Sref{sugra2c}, and violate the global $\bl$ symmetry
which is valid at the level of $W$.

\subsection{Canonical Normalization}\label{ci1}

The first step towards the establishment of CSI is the canonical
normalization of the fields involved in the parametrization of
$\phc$ in \Eref{para}. This can be done if we identify the kinetic
term on \Eref{action1} with the canonical ones as follows
\beq \label{kin} \vevi{K_{\Phi\Phi^*}}|\dot \Phi|^2
\simeq\frac12\lf\dot{\what\phi}^{2}+\dot{\what
\theta}^{2}\rg~~\Rightarrow~~\frac{d\se}{d\sg}=J={\sqrt{2N}\over\fp}~~~\mbox{and}~~~
\widehat{\theta}\simeq
J\sg\theta~~~\mbox{with}~~~\vevi{K_{\Phi\Phi^*}}=\frac{N}{\fp^2},
\eeq
found from \Eref{ma} if we restrict our attention on the path in
\Eref{inftr}. Upon integrating the $\sg-\se$ relation above we
arrive at \Eref{tmd} in accordance with our aim.

\subsection{Inflationary Potential}\label{ci2}

The second step in our attempt to implement SI is the reproduction
of $\Vhi$ of \Eref{vhi} starting from \Eref{Vhio} with $W$ given
in \Eref{wci} and $K=\tkbad$ in \Eref{tkbad}. This is trivially
verified if we take into account the field configuration of
\Eref{inftr}. However, $\Vhi$ of \Eref{vhi} is a tree-level result
which may receive radiative corrections. These can be estimated by
constructing the mass spectrum of the theory along the trajectory
in \Eref{inftr}. Our results are summarized in \Tref{tab3}, where
we arrange the expressions of the masses squared $\what
m^2_{\chi^\al}$ (with $\chi^\al=\theta$ and $s$) divided by
$\Hhi^2\simeq\Vhi/3$. From them we can appreciate the role of
$N_S<6$ in retaining positive $\what m^2_{s}$ and thereby
stabilizing the track in \Eref{inftr}. Also, we confirm that
$\what m^2_{\chi^\al}\gg\Hhi^2$ for $\sg\leq1$ and so we do not
obtain inflationary primordial perturbation from other fields
besides $\sg$. In \Tref{tab3} we display also the masses $\what
m^2_{\psi\pm}$ of the corresponding fermions too -- we define
$\what\psi_{\Phi}=J\psi_{\Phi}$ where $\psi_\Phi$ and $\psi_S$ are
the Weyl spinors associated with $S$ and $\Phi$ respectively.
Considering SUGRA as an effective theory with cutoff scale equal
to $\mP$, the well-known Coleman-Weinberg formula -- see e.g
\cref{jhep} -- can be employed self-consistently taking into
account the masses which lie well below $\mP$, i.e., all the
masses arranged in \Tref{tab3}. Therefore, the one-loop correction
to $\Vhi$ reads
\beq\dV={1\over64\pi^2}\lf \widehat m_{\theta}^4\ln{\widehat
m_{\theta}^2\over\Lambda^2} +2
m_{s}^4\ln{m_{s}^2\over\Lambda^2}-4\widehat
m_{\psi_{\pm}}^4\ln{\widehat m_{\psi_{\pm}}^2\over\Lambda^2}\rg
,\label{Vhic}\eeq
where $\Lambda$ is a \emph{renormalization group} ({\sf\small RG})
mass scale. The resulting $\dV$ lets intact our inflationary
outputs, provided that $\Lambda$ is determined by requiring
$\dV(\sgx)=0$ or $\dV(\sgf)=0$. Namely, solving these conditions
w.r.t $\Lambda$ we obtain
\beq \label{Ld} \Lambda=e^{{c_m}/{c_{\Lambda}}}~~\mbox{with}~~\bcs
c_m=\widehat m_{\theta}^4\ln{\widehat m_{\theta}^2} +2
m_{s}^4\ln{m_{s}^2}-4\widehat
m_{\psi_{\pm}}^4\ln{\widehat m_{\psi_{\pm}}^2}, \\
c_{\Lambda}= \widehat m_{\theta}^4+2 m_{s}^4-4\widehat
m_{\psi_{\pm}}^4. \ecs\eeq
If determined for $\sg=\sgx$ or $\sg=\sgf$, the expression above
yields $\Lambda_\star=\Lambda(\sgx)$ or $\Lambda_{\rm
f}=\Lambda(\sgf)$. Both choices let intact the inflationary
observables derived exclusively by using the (tree level) $\Vhi$
in \Eref{vhi} as shown for a similar model in \cref{jhep}.
Moreover, the renormalization-group running is expected to be
negligible because $\Ld$ is close to the inflationary scale $\Hhi$
-- see below.

\renewcommand{\arraystretch}{1.4}
\begin{table}[!t]
{\small \bec\begin{tabular}{|c||c||c|l|l|}\hline {\sc Fields}&{\sc
Eigenstates}& \multicolumn{3}{|c|}{\sc Masses Squared$/\Hhi^2$}\\
\hline\hline $1$ real scalar &$\widehat \theta$ & $\widehat
m^2_{\theta}/\Hhi^2$& \multicolumn{2}{|c|}{$
3(\nd(1-\sg)^2+4N\sg)/2N\sg\simeq6$}\\
$2$ real scalars &$s,~\bar s$ & $m^2_{
s}/\Hhi^2$&\multicolumn{2}{|c|}{$6/N_S+3\nd^2(1-\sg)^2(n\fd-\nd\sg)^2/4N\sg^2$}\\\hline
$2$ Weyl spinors & ${(\what{\psi}_{\Phi}\pm
\what{\psi}_{S})/\sqrt{2}}$& $\what m^2_{ \psi\pm}/\Hhi^2$&
\multicolumn{2}{|c|}{$ 3(1-\sg)^2(n\fd-\nd\sg)^2/8N\sg^2$}
\\ \hline
\end{tabular}\eec}
\caption{\sl\small Mass spectrum for CSI with $K=\tkbad$ along the
inflationary trajectory of \Eref{inftr}.}\label{tab3}
\end{table}\renewcommand{\arraystretch}{1}

\section{Gauge non-Singlet Inflaton}\label{hi}

In the present scheme the inflaton field can be identified by the
radial component of a conjugate pair of Higgs superfields. We here
focus on the Higgs superfields, $\bar\Phi$ and $\Phi$ which break
the GUT symmetry $\Ggut=\Gsm\times\bl$ down to SM gauge group
$\Gsm$ through their v.e.vs. We parameterize the involved
superfields as follows
\beq \Phi=\phi e^{i\theta}\cos\theta_\Phi\>\>\>
\mbox{and}\>\>\>\bar\Phi=\phi
e^{i\bar{\theta}}\sin\theta_\Phi\>\>\>
\mbox{with}\>\>\>0\leq\theta_{\Phi}\leq{\pi}/{2}~~~\mbox{and}~~~S=(s+i\bar
s)/\sqrt{2}.\label{hpara}\eeq
Note that superfield $S$ is $\Ggut$ singlet. As we verify in
\Sref{hi2} below, a D-flat direction is
\beq
\vevi{\theta}=\vevi{\bar{\theta}}=0,\>\vevi{\theta_{\Phi}}={\pi/4}\>\>\>\mbox{and}\>\>\>\vevi{S}=0,\label{inftrh}\eeq
which can be qualified as inflationary path. We below outline the
SUGRA setting in \Sref{hi0} and determine the inflationary
potential in \Sref{hi2} after canonically normalize the various
fields in \Sref{hi1}. Finally, we give some informations for the
$\bl$ phase transition in \Sref{hi3}.

\subsection{Set-up}\label{hi0}

In accordance with our discussion in \Sref{sugra2b}, we consider
$\ffw$ as a function of the $\Ggut$-invariant holomorphic quantity
$\bar\Phi\Phi$, i.e.,
\beq \ffw=(2\bar\Phi\Phi)^{n/4}-M^2 ~~~\mbox{and so}~~~W= \ld S\lf
(2\bar\Phi\Phi)^{n/4}-M^2\rg,\label{whi} \eeq
where $\ld$ and $M\ll1$ are free parameters whereas $n$ is a
multiplier of $4$. $W$ is determined for $n=4$ if we impose an $R$
symmetry under which $W$ has the charge of $S$ whereas the
combination $\bar\Phi\Phi$ is uncharged.  These two symmetries do
not disallow, however, terms of the form $(\phcb\phc)^p$ with
$p>2$ in $W$ and so stabilization of SI against corrections from
those $W$ terms dictates \sub\ values for $\bar\Phi$ and $\Phi$
or, via \Eref{hpara}, $\sg$. On the other hand, for $n>4$ the
determination of $W$ uniquely requires the imposition of an extra
discrete symmetry $\mathbb{Z}_{n/4}$ under which $\bar\Phi\Phi$
has unit charge.

The realization of HSI can be accomplished by the consideration of
two possible $K$'s which are consistent with the imposed
symmetries. They incorporate the following functions
\beq
\ffe=1-2\phcb\phc,~\ffd=1+\sqrt{2\phcb\phc}~~~\mbox{and}~~~\ffm=
\bcs \lf(1-2|\phc|^2)(1-2|\phcb|^2)\rg^{1/2}&\mbox{for}~~K=\tkaa,\\
1-|\phc|^2-|\phcb|^2&\mbox{for}~~K=\tkbah,\ecs \label{ffs}\eeq
where the involved $K$'s have been first introduced in \cref{sor}
and read
\beq\tkaa=-\frac{\nm}{2}\ln\frac{(1-2|\phc|^2)(1-2|\phcb|^2)}
{(1-2\phcb\phc)(1-2\phcb^*\phc^*)}~~~\mbox{and}~~~\tkbah=-\nm\ln\frac{1-|\phc|^2-|\phcb|^2}
{\lf(1-2\phcb\phc)(1-2\phcb^*\phc^*)\rg^{1/2}}. \label{tkhi}\eeq
These $K$'s enjoy a shift symmetry, as shown in Appendix
\ref{app}, whose the violation is expressed by
\beq K_{\rm d}=
-\frac{\nd}{2}\ln(1+\sqrt{2\phcb\phc})-\frac{\nd}{2}\ln(1+\sqrt{2\phcb^*\phc^*}).\label{kdh}\eeq
Therefore, the total $K$'s for the two versions of HSI considered
here are
\beq
\tkbaad=\kb+\tkaa+\kd~~~\mbox{and}~~~\tkbbad=\kb+\tkbah+\kd,\label{khi}\eeq
where $\kb$ is given by \Eref{kb}. These $K$'s parameterize
\cite{sor} respectively the \Km s
\beqs\beq \label{mnfh} \mnfbaa=\lf\frac{SU(2)}{U(1)}\rg_S\times\lf
\frac{SU(1,1)}{U(1)}\rg^2_{\phcb\phc} ~~~\mbox{or}~~~\mnfbba=
\lf\frac{SU(2)}{U(1)}\rg_S \times\lf\frac{SU(2,1)}{SU(2) \times
U(1)}\rg_{\phcb\phc}\,, \eeq
with moduli-space scalar curvatures correspondingly \cite{sor}
\beq \label{Rmnf} {\sf
R}_{2(11)^2}=2/\nus-{8}/{N}\>\>\>\mbox{and}\>\>\>{\sf
R}_{221}=2/\nus-{6}/{N}.\eeq\eeqs
Note that we apply in \Eref{mnfh} the same notation for the
indices as in \Eref{mnfs}.

\subsection{Canonical Normalization}\label{hi1}

To obtain SI we have to correctly identify the canonically
normalized (hatted) fields of the $\phcb-\phc$ system, defined as
follows
\beq \vevi{K_{\al\bbet}}\dot z^\al \dot z^{*\bbet} \simeq
\frac12\lf\dot{\widehat \sg}^2+\dot{\widehat
\theta}_+^2+\dot{\widehat \theta}_-^2+\dot{\widehat
\theta}_\Phi^2\rg, \label{kinh}\eeq
where the elements $\vevi{K_{\al\bbet}}$ for the $K$'s in
Eq.~(\ref{tkhi}) are contained in the matrix ${\bf M}_{\phcb\phc}$
-- see \Eref{me} of Appendix~\ref{app} -- whose the form in the
limit of \Eref{inftrh} is
\beq \vevi{{\bf M}_{\phcb\phc}}=\begin{cases}
\kp\,\diag\lf1,1\rg&\mbox{for}\>\>\>K=\tkaa, \\
\kp\mtta{1-\sg^2/2}{\sg^2/2}{\sg^2/2}{1-\sg^2/2}\>\>\>&\mbox{for}\>\>\>K=\tkbah,
\end{cases}~~
\mbox{with}~~~\kp=N/\fp^{2}. \label{Mk}\eeq
Expanding the second term of the right-hand side of \Eref{action1}
along the path in \Eref{inftr} for $\al=\phcb,\phc$ and
substituting there \Eref{Mk}, we obtain
\beqs\beq \label{kzz} \vevi{K_{\al\bbet}}\dot z^\al \dot
z^{*\bbet}=\begin{cases}{\kp}\dot
\sg^2+{\kp}\sg^2\lf\dot\theta^2_+
+\dot\theta^2_-+2\dot\theta^2_\Phi\rg/2&\mbox{for}~~~K=\tkaa,
\\ {\kp_+}\lf\dot \sg^2+\sg^2\dot\theta^2_+/2
\rg+{\kp_-\sg^2}\lf\dot\theta^2_-/2 +\dot\theta^2_\Phi
\rg&\mbox{for}~~~K=\tkbah,
\end{cases}\eeq 
with $\kp_+=\kp$, $\kp_-=\kp\fp$ and
$\theta_{\pm}=\lf\bar\theta\pm\theta\rg/\sqrt{2}$. Comparing
\eqs{kzz}{kinh} we can derive the relation between the hatted and
unhatted fields. As regards the inflaton, the equality between
$\kp$ and $\kp_+$ in \Eref{kzz} assures that, for both $K$'s, the
$d\se/d\sg$ relation is identical with that found in \Eref{kin}
and so the correct $\sg-\se$ relation in \Eref{tmd} emerges. For
the remaining fields of the $\phcb-\phc$ system we find
\beq \label{Je1} \>\>\>\>\>\>\>\>\>\>\>\bem
 \begin{array}{ll}\widehat{\theta}_\pm
={\sqrt{\kp}}\sg\theta_\pm,\>\>\widehat \theta_\Phi =
\sqrt{2\kp}\sg\lf\theta_\Phi-{\pi}/{4}\rg
\>\>\>&\mbox{for}\>\>\>K=\tkaa\,,\\ \widehat{\theta}_+
=\sqrt{\kpp}\sg\theta_+,\>\>\widehat{\theta}_-
=\sqrt{{\kpm}}\sg\theta_-,\>\>\widehat \theta_\Phi =
\sqrt{2\kpm}\sg\lf\theta_\Phi-{\pi}/{4}\rg
\>\>\>&\mbox{for}\>\>\>K=\tkbah,\end{array}\eem\eeq\eeqs
where we take into account that the masses of the scalars besides
$\se$ during SI are large enough such that the dependence of the
hatted fields on $\sg$ does not influence their dynamics. Needless
to say, the extra contributions into the $K$'s in \Eref{tkhi} do
not disturb our formulae above.

\subsection{Inflationary Potential}\label{hi2}

Upon substitution of $W$ and $K$ from \eqs{whi}{tkhi} into
\Eref{Vhio} we arrive at the advertised form of $\Vhi$ in
\Eref{vhi}. As regards $V_{\rm D}$ -- see \Eref{Vsugra} --, for
the $K$'s in \Eref{tkhi}, ${\rm D}_{X}$ takes the form
\beq \label{Dbl} {\rm D}_{X}=
N\lf|\phc|^2-|\phcb|^2\rg\cdot\begin{cases}
%
(1-2|\phcb|^2)^{-1}(1-2|\phcb|^2)^{-1}&\mbox{for}\>\>\>K=\tkbaad,\\
\lf1-|\phc|^2-|\phcb|^2\rg^{-1}&\mbox{for}\>\>\>K=\tkbbad\,.\end{cases}\eeq
If we insert this result in \Eref{Vsugra} and take the limit of
\Eref{inftrh} we deduce that $\vevi{V_{\rm D}}=0$ and so no D-term
contribution arises in $V_{\rm SUGRA}$ during HSI.

We can also proceed in deriving the mass spectrum of the models
along the direction of \Eref{inftrh} and verifying its stability
against the fluctuations of the non-inflaton fields. The results
of our computation are accumulated in \Tref{tab1}.  As for
spectrum in \Tref{tab3}, $N_S<6$ plays a crucial role in retaining
positive and heavy enough $\what m^2_{s}$ whereas
$\widehat\theta_{+}$ turns out to be spontaneously heavy enough as
$\widehat\theta$ in \Tref{tab1}. Here, however, we also display
the masses, of $\widehat \theta_\Phi$, of the gauge boson $A_{X}$
and of the corresponding fermions which acquire contribution from
the gauge sector of the theory and so they are safely heavy and
stabilized. The unspecified eigenstate $\what \psi_\pm$ is defined
as
\beq \what \psi_\pm=(\what{\psi}_{\Phi+}\pm
{\psi}_{S})/\sqrt{2}\>\>\>\mbox{where}\>\>\>\psi_{\Phi\pm}=(\psi_{\bar\Phi}\pm\psi_\Phi)/\sqrt{2}\,\eeq
with the spinors $\psi_S$ and $\psi_{\Phi\pm}$ being associated
with the superfields $S$ and $\bar\Phi-\Phi$. Comparing the
various masses we notice only minor discriminations between the
two analyzed $K$'s.

The non-zero $M_{X}$ signals the fact that $\bl$ is broken during
SI since $A_{X}$ becomes massive absorbing the massless Goldstone
boson associated with $\what\theta_-$. As a consequence, six
degrees of freedom before the spontaneous breaking (four
corresponding to the two complex scalars $\phcb$ and $\phc$ and
two corresponding to the massless gauge boson $A_X$ of $U(1)_X$)
are redistributed as follows: three degrees of freedom are
associated with the real propagating scalars ($\what\theta_+$,
$\what\theta_\phc$ and $\what\sg$), whereas the residual one
degree of freedom combines together with the two ones of the
initially massless gauge boson $A_X$ to make it massive. From
\Tref{tab1}, we can also deduce that the numbers of bosonic
(eight) and fermionic (eight) degrees of freedom are equal if we
take into account the inflaton $\sg$ not included.

The derived mass spectrum allows us to determine the one-loop
radiative corrections to $\Vhi$ employing the Coleman-Weinberg
formula -- see e.g. \cref{jhep}. However, we remark that
$M_{X}^2\gg\mP^2$ and $\what m_{\theta_\Phi}^2\gg\mP^2$ and so
these masses can not be included in the formula above \cite{jhep}.
As a consequence, $\dV$ and $\Ldx$ assume the same expressions as
in \eqs{Vhic}{Ld} respectively with the relevant masses replaced
by those defined in \Tref{tab1}.


\renewcommand{\arraystretch}{1.4}
\begin{table}[!t]
\begin{center}
{\small \begin{tabular}{|c||c|c||c|c|}\hline {\sc Fields}&{\sc
Eigen-}& \multicolumn{3}{|c|}{\sc Masses Squared}\\\cline{3-5}
&{\sc states}& &
\multicolumn{1}{|c|}{$K=\kbaad$}&\multicolumn{1}{|c|}{$K=\kbbad$}\\
\hline\hline
%
4 &$\widehat\theta_{+}$&$m_{\widehat\theta+}^2$&
\multicolumn{2}{|c|}{$3\Hhi^2(\nd(1-\sg)^2+4N\sg)/4N\sg\simeq3\Hhi^2$}\\\cline{4-5}
real&$\widehat \theta_\Phi$ &$\widehat m_{
\theta_\Phi}^2$&\multicolumn{1}{|c|}{$M^2_{X}+6\Hhi^2-3\Hhi^2\cdot$}&
{$M^2_{X}+6\Hhi^2-3\Hhi^2\cdot$}\\
scalars&&&\multicolumn{1}{|c|}{$(1-\sg)^2\fd(n\fd-\nd\sg)/N\sg^2$}&
{$(1-\sg)(n\fd-\nd\sg)/N\sg^2$}\\ \cline{4-5}
&$s, {\bar{s}}$ &$ \widehat m_{
s}^2$&\multicolumn{2}{|c|}{$\Hhi^2(6/\nus+3(1-\sg)^2(n\fd-\nd\sg)^2/N\sg^2)$}\\
\hline
1 gauge boson &{$A_{X}$}&{$M_{X}^2$}&
\multicolumn{1}{|c|}{$2Ng^2\sg^2/\fm^2$}&$2Ng^2\sg^2/\fm$\\\hline
$4$ Weyl  & $\what \psi_\pm$ & $\what m^2_{ \psi\pm}$&
\multicolumn{2}{|c|}{$3(1-\sg)^2(n\fd-\nd\sg)^2\Hhi^2/8N\sg^2$}\\\cline{4-5}
spinors&$\ldu_{X}, \widehat\psi_{\Phi-}$&$M_{X}^2$&\multicolumn{1}{|c|}{$2Ng^2\sg^2/\fp^2$}&$2Ng^2\sg^2/\fp$\\
\hline
\end{tabular}}\end{center}
\caption{\sl\small Mass spectrum for HSI with $K=\tkbaad$ and
$\tkbbad$ along the inflationary trajectory of \Eref{inftrh}.
}\label{tab1}
\end{table}\renewcommand{\arraystretch}{1.}

\subsection{$\bl$ Phase Transition}\label{hi3}

In the context of HSI, $W$ in \Eref{whi} leads not only to an
inflationary era but also to the breaking of $\bl$ symmetry. In
our introductory set-up the v.e.vs of $\phcb$ and $\phc$ break
$\bl$ down to $\mathbb{Z}^{X}_2$. Indeed, minimizing $\Vhi$ in
\Eref{vhi} with $\sg\ll\mP$ we find that a SUSY vacuum arises
after the end of HSI determined as follows
\beq \vev{S}=0 \>\>\>\mbox{and}\>\>\> |\vev{2\bar\Phi\Phi}|^{n/4}=
M^2~\Rightarrow~\vev{\sg}=M^{4/n}. \label{vevs} \eeq
Although $\vev{\Phi}$ and $\vev{\bar\Phi}$ break spontaneously
$\bl$, no cosmic strings  are produced at the SUSY vacuum, since
this symmetry is already broken during HSI -- cf.~\cref{jhepcs}
The contributions from the soft SUSY breaking terms can be safely
neglected within contemporary SUSY, since the corresponding mass
scale is much smaller than $M$. They may shift \cite{R2r,
unh,ighi,sor2,jhepcs}, however, slightly $\vev{S}$ from zero in
\Eref{vevs}.

As regards the value of $M$, it can be determined by requiring
that $\vev{\bar\Phi\Phi}$ takes the value dictated by the
unification of MSSM gauge coupling constants. In particular, the
unification scale
$\mgut\simeq2/2.433\times10^{-2}\simeq8.22\cdot10^{-3}$ is to be
identified with $\vev{M_{X}}$ -- see \Tref{tab1} --, i.e.,
\beq \label{Mg} \vev{M_{X}}=\sqrt{2N}gM\simeq\mgut
\>\>\Rightarrow\>\>M\simeq\lf{\mgut}/{g\sqrt{2N}}\rg^{n/4}
\>\>\>\mbox{for}\>\>\>\vev{\fp}\simeq1. \eeq
Here $g\simeq0.7$ is the value of the GUT gauge coupling constant.

The determination of $M$ influences heavily the inflaton mass at
the vacuum, $\msn$ and induces an $N$ dependence in the results.
Indeed, the (canonically normalized) inflaton,
\beq\dphi=\vev{J}\dph\>\>\>\mbox{with}\>\>\> \dph=\phi-\vev{\sg}
\>\>\>\mbox{and}\>\>\>\vev{J}=\sqrt{2N}/\vev{\fp}\label{dphi} \eeq
acquires mass, at the SUSY vacuum in \Eref{vevs}, which is given
by
\beq \label{msn} \msn=\left\langle\Ve_{\rm
I,\se\se}\right\rangle^{1/2}= \left\langle \Ve_{\rm
I,\sg\sg}/J^2\right\rangle^{1/2}=\frac{\ld n M^{2(1 - 2/n)}}{2
\sqrt{\nm}}\frac{1- M^{8/n}}{\lf 1 + M^{4/n}\rg^{\nd/2}}.\eeq
Since $M\ll\mP$, this result is essentially valid for both $K$'s
in \Eref{tkhi}. Note in passing that the mass of the inflaton for
CSI with $n=2$ is given by $\msn=\sqrt{2/\nm}\ld\mP$.

\section{Inflation Analysis}\label{fti}

We proceed now to the analytic and numeric investigation of the
viability of our models in \Srefs{hiobs} and \ref{num}
respectively.

\subsection{Analytic Results}\label{hiobs}

Since both models, CSI and HSI, are based on the same $\sg-\se$
relation in \Eref{tmd} and the same $\Vhi$ in \Eref{vhi}, their
analysis can be performed in a unified way. Namely, the period of
slow-roll SI is determined by the condition -- see, e.g.,
\cref{rev, rev1}:
\beqs\beq{\ftn\sf
max}\left\{\eph(\se),\left|\ith(\se)\right|\right\}\simeq1,\>\mbox{where}\>\>
\eph=\frac12\left(\frac{\Ve_{\rm I,\se}}{\Ve_{\rm
I}}\right)^2\>\>\>\mbox{and}\>\>\> \ith={\Ve_{\rm
I,\se\se}\over\Ve_{\rm I}} \label{srcon} \eeq
are the slow-roll parameters which and can be estimated employing
$J$ in \Eref{kin} without express explicitly $\Vci$ in terms of
$\se$. Indeed, our results are found to be
\bea\nonumber &&~~~~~~~~~~~~~~~~~~~~~~~\epsilon= \frac{(\sg-1)^2
(n\fd-\nd \sg)^2}{2 N
\sg^2}~~~\mbox{and}~~~ \eta=\frac{(\sg-1)}{\nm\sg^2}\cdot\\
&&\left(n^2 (\sg-1)\fd^2+n \left((1-2 \nd) \sg^3+2 \nd
\sg+\sg^2+\fd\right)+\nd \sg^2 (\nd (\sg-1)-\fd)\right).
\label{et}\eea\eeqs
\Eref{srcon} is saturated for $\sg=\sgf$, which is the maximal
values from the following two solutions
\beqs\bea \sg_{1\rm f}&\simeq& \frac{\sqrt{4. n^2+2 n \nm-4 n
\nd+\nd^2}-\nd}{2 (2 n+\nm-2 \nd)}; \label{sgf1}\\
\sg_{2\rm f}&\simeq& \frac{\sqrt{4 n^2 \nd^2-4n \left(n-1\right)
\left(-n^2+2 n \nd+n+\nm\right)}+2n \nd}{2 \left(2
n-n^2+\nd+n+\nm\right)}\cdot\eea\eeqs
In practice, for $\nd<n/2$, $\sg_{1\rm f}<\sg_{2\rm f}$ and so SI
terminates at $\sgf=\sg_{2\rm f}$ whereas larger $\nd$'s give rise
to the inverse hierarchy and so $\sgf=\sg_{1\rm f}$.

The number of e-foldings $\Ns$ that the pivot scale $\ks=0.05/{\rm
Mpc}$ experiences during SI is estimated as
\begin{equation}
\Ns=\int_{\sef}^{\sex} d\se\frac{\Vci}{\Ve_{\rm I,\se}}=N\lf
I_N(\sgx)-I_N(\sgf)\rg, \label{Nhi}\eeq
where $\sex$ and $\sgx$ are the value of $\se$ and $\sg$
respectively when $\ks$ crosses outside the inflationary horizon
and the involved function $I_N$ reads
\begin{equation}
I_N(\sg)= \frac{\nd}{4 \dn^2}\ln(1 - \sg)-\frac{1}{2 \dn (1-\sg)}
- \frac{1}{4 \nd}\ln\fd + \frac{n(n-\nd)}{\nd\dn^2}\ln(n\fp
-\nd\sg),\label{In}\eeq
where $\dn=\nd-2n<0$. The last inequality stems from the fact that
the dominant contribution to $I_N$ originates from the
non-logarithmic term with $\sg<1$. The presence of a pole in $J$
-- see \Eref{kin} -- and the effective nature of SUGRA forces us
to work in this domain of $\sg$. Consequently, the positivity of
$\Ns$ implies the upper bound above on $\dn$ which restricts
seriously the allowed region of our model as we see in \Sref{num}.

Due to the complicate form of $\Ns$ in \Eref{Nhi}, it is not
doable to solve the equation above w.r.t $\sgx$ and find a generic
analytical expression for it and the inflationary observables --
see below. As a consequence, our last resort is the numerical
computation, whose the results are presented in \Sref{num}.
Nonetheless, for $\nd\ll n$ and taking into account $\sgx\gg\sgf$,
we can derive an approximate and rather accurate formula for $\Ns$
since it is dominated from the non-logarithmic term of $I_N$. In
this portion of parameter space we can determine $\sgx$ as follows
\begin{equation}
\label{Nhi1} \Ns\simeq -\frac{\nm}{2\dn}
\frac{\sgx}{1-\sgx}~~\Rightarrow~~\sgx\simeq\frac{2 \dn\Ns}{2
\dn\Ns - \nm}.
\end{equation}
Since both factors of the ratio above are negative and the
denominator is larger in absolute value we expect that $\sgx<1$.
Therefore, our proposal can be stabilized against corrections from
higher order terms in the $K$'s -- see \Sref{hi1}.

The amplitude $\As$ of the power spectrum of the curvature
perturbations generated by $\sg$ can be computed using the
standard formulae
\begin{equation}
\label{As} \As^{1/2}= \frac{1}{2\sqrt{3}\, \pi} \; \frac{\Ve_{\rm
I}^{3/2}(\sex)}{|\Ve_{\rm
I,\se}(\sex)|}=\frac{\ld\sgx\sqrt{\nm}\sqrt{\sgx^n (1 +
\sgx)^{-\nd}}}{2 \sqrt{3}\pi(1 - \sgx) (n\fps - \nd\sgx)},\eeq
where $\fps=\fp(\sgx)$. From the right formula in \Eref{As} we can
derive a relation between $\ld$ and $\As$. For simplicitly we set
$\nd=0$ and so we find
\begin{equation}
\label{lan} \ld\simeq\frac{\pi n^{-\frac{n}{2}}\sqrt{3\As}
\sqrt{\nm}   (8 n \Ns+\nm)}{2^{n+1}\Ns^{\frac{n}{2}+1}(4 n
\Ns+\nm)^{1-{n/2}}}.
\end{equation}
It is clear that  no $\dn$  depedence appears due to drastic
simplification done. However, the numerical result is quite close
to the exact one given that $\nd$ is bounded above as noticed from
\Eref{In}.

The remaining inflationary observables -- i.e., the (scalar)
spectral index $n_{\rm s}$, its running $a_{\rm s}$, and the
scalar-to-tensor ratio $r$ -- are found from the relations
\cite{rev,rev1}
\beq \label{nras} \ns=1-6\eph_\star\ +\
2\ith_\star,~r=16\eph_\star~~~\mbox{and}~~~\as
=\:2\left(4\ith_\star^2-(n_{\rm
s}-1)^2\right)/3-2\widehat\xi_\star, \eeq
where the variables with subscript $\star$ are evaluated at
$\sg=\sgx$ and $\widehat\xi={\Ve_{\rm I,\widehat\phi} \Ve_{\rm
I,\widehat\phi\widehat\phi\widehat\phi}/\Ve^2_{\rm I}}$. Inserting
$\sgx$ from \Eref{Nhi1} into \Eref{et} and then into equations
above we can obtain the following approximate expressions
\beqs\bea  \ns&\simeq&1-\frac{2}{\Ns}-\frac{\nd^2 \nm}{4 \Ns^2
\dn^2}-\frac{n^2 \nm}{\Ns^2 \dn^2}+\frac{\nd n \nm}{\Ns^2
\dn^2}-\frac{3 \nd \nm}{2 \Ns^2 \dn^2}+\frac{n \nm}{\Ns^2 \dn^2},\label{nsa}\\
r&\simeq&\frac{2 \nm}{\Ns^2}+\frac{2 \nd \nm^2}{\Ns^3 \dn^2}- \frac{2 n \nm^2}{\Ns^3 \dn^2},\label{rsa} \\
\as&\simeq&-\frac{2}{\Ns^2}-\frac{\nd^2 \nm}{2 \Ns^3
\dn^2}-\frac{2 n^2 \nm}{\Ns^3 \dn^2}+\frac{2 \nd n \nm}{\Ns^3
\dn^2}-\frac{7 \nd \nm}{2 \Ns^3 \dn^2}+\frac{2 n \nm}{\Ns^3\dn^2}.
\label{asa}\eea\eeqs
These expressions give accurate results for $\nd\ll n$ or
$\dn\simeq -2n$. For $n=\nd=2$ the above results converge to those
obtained for the pure $\alpha$-SI
\cite{tkrefa,eno7,ellis21,linde21}, i.e.,
\beq (\ns,r,\as)\simeq(1-2/\Ns,2N/\Ns^2,-2/\Ns^2).
\label{attr}\eeq
The same results are obtained (for reasonably low $n$ and $N$
values) in the limit $\nd=0$ where the pure T-model inflation is
revealed.

\subsection{Numerical Results}\label{num}

Our estimations above can be verified and extended for any $\dn$
numerically. In particular, we confront the quantities in
\eqs{Nhi}{As} with the observational requirements \cite{plcp}
\beq\Ns \simeq61.3+\frac{1-3w_{\rm rh}}{12(1+w_{\rm
rh})}\ln\frac{\pi^2g_{\rm rh*}\Trh^4}{30\Vci(\sgf)}+
\frac12\ln\lf{\Vci(\sgx)\over g_{\rm
rh*}^{1/6}\Vci(\sgf)^{1/2}}\rg~~~
\mbox{and}~~~\As^{1/2}\simeq4.588\cdot10^{-5},\label{prob}\eeq
where we assume that SI is followed in turn by an oscillatory
phase with mean equation-of-state parameter $w_{\rm rh}$,
radiation and matter domination. Motivated by implementations
\cite{R2r, ighi, blst, sor2} of non-thermal leptogenesis, which
may follow SI, we set $\Trh\simeq1~\EeV$ for the reheat
temperature. Also, we take for the energy-density effective number
of degrees of freedom $g_{\rm rh*}=228.75$ which corresponds to
the MSSM spectrum. Note that this $\Trh$ avoids exhaustive tuning
on the relevant coupling constant involved in the decay width of
the inflaton -- cf. \cref{phenoAt,ellis21}.

Due to the peculiar expression of $\Vhi$ in \Eref{vhi} and the
non-minimal kinetic mixing in \Eref{tmd}, the estimation of $\wrh$
requires some care -- cf. \cref{wreh,wreha,wtmd}. We determine it
adapting the general formula \cite{turner}, i.e.
\beq w_{\rm rh}=2\frac{\int_{\sgn}^{\sgm} d\sg J(1-
\Vhi/\Vhi(\sgm))^{1/2}}{\int_{\sgn}^{\sgm} d\sg J(1-
\Vhi/\Vhi(\sgm))^{-1/2}}-1,\label{wrh}\eeq
where $\sgn=0$ for CSI whereas $\sgn=\vev{\sg}$ given in
\Eref{vevs} for HSI. The amplitude of the oscillations during
reheating $\sgm$ is found by solving numerically the condition
$\sqrt{3}\Hhi(\sgm)=\msn$ if $\msn<\sqrt{3}\Hhi(\sgf)$ or it is
$\sgm=\sgf$ otherwise. The result deviates slightly from the naive
expectation according to which
\beq w_{\rm rh}=(n-2)/(n+2),\label{wrha}\eeq
for a monomial power-law potential of the form $\sg^n$.

Enforcing \Eref{prob} we can restrict $\ld$ and $\sgx$ via
\Eref{Nhi}. In general, we obtain $\ld\sim10^{-5}$ in agreement
with \Eref{lan}. Regarding $\sgx$ we assume that $\sg$ starts its
slow roll below the location of kinetic pole, i.e., $\sg=1$,
consistently with our approach to SUGRA as an effective theory
below $\mP=1$. The closer to pole $\sgx$ is set the larger $\Ns$
is obtained. Therefore, a tuning of the initial conditions is
required which can be somehow quantified defining the quantity
\beq \Dex=\left(1-\sgx\right).\label{dex}\eeq
The naturalness of the attainment of SI increases with $\Dex$.
After the extraction of $\ld$ and $\sgx$, we compute the models'
predictions via \Eref{nras}, for any selected values for the
remaining parameters, $N$, $n$ and $\nd$ -- with $M\ll1$. Our
outputs are encoded as lines in the $\ns-r$ plane and compared
against the observational data \cite{gws}. We take into account
the latest \emph{Planck release 4} ({\sf\small PR4}) -- including
TT,TE,EE+lowE power spectra \cite{gws1} --, \emph{Baryon Acoustic
Oscillations} ({\sf\small BAO}), CMB-lensing and BICEP/{\it Keck}
({\sf\small BK18}) data. Fitting it \cite{gws} with $\Lambda$CDM
we obtain the marginalized joint $68\%$ [$95\%$] regions depicted
by the dark [light] shaded contours in the aforementioned figures.
Approximately we obtain
\beq \label{nspl} {\sf\small
(a)}~~\ns=0.965\pm0.009~~~\mbox{and}~~~{\sf\small
(b)}~~r\lesssim0.032 \eeq
at 95$\%$ \emph{confidence level} ({\sf\small c.l.}) with
negligible $\as$ -- cf. \cref{ellis21}. The results are exposed
separately, in \Sref{res2a} and \ref{res2b} for CSI and HSI
respectively.

\subsubsection{SI with a Gauge-Singlet Inflaton (CSI).}\label{res2a}

\begin{figure}[!t]\vspace*{-.12in}
\hspace*{-.19in}
\begin{minipage}{8in}
\epsfig{file=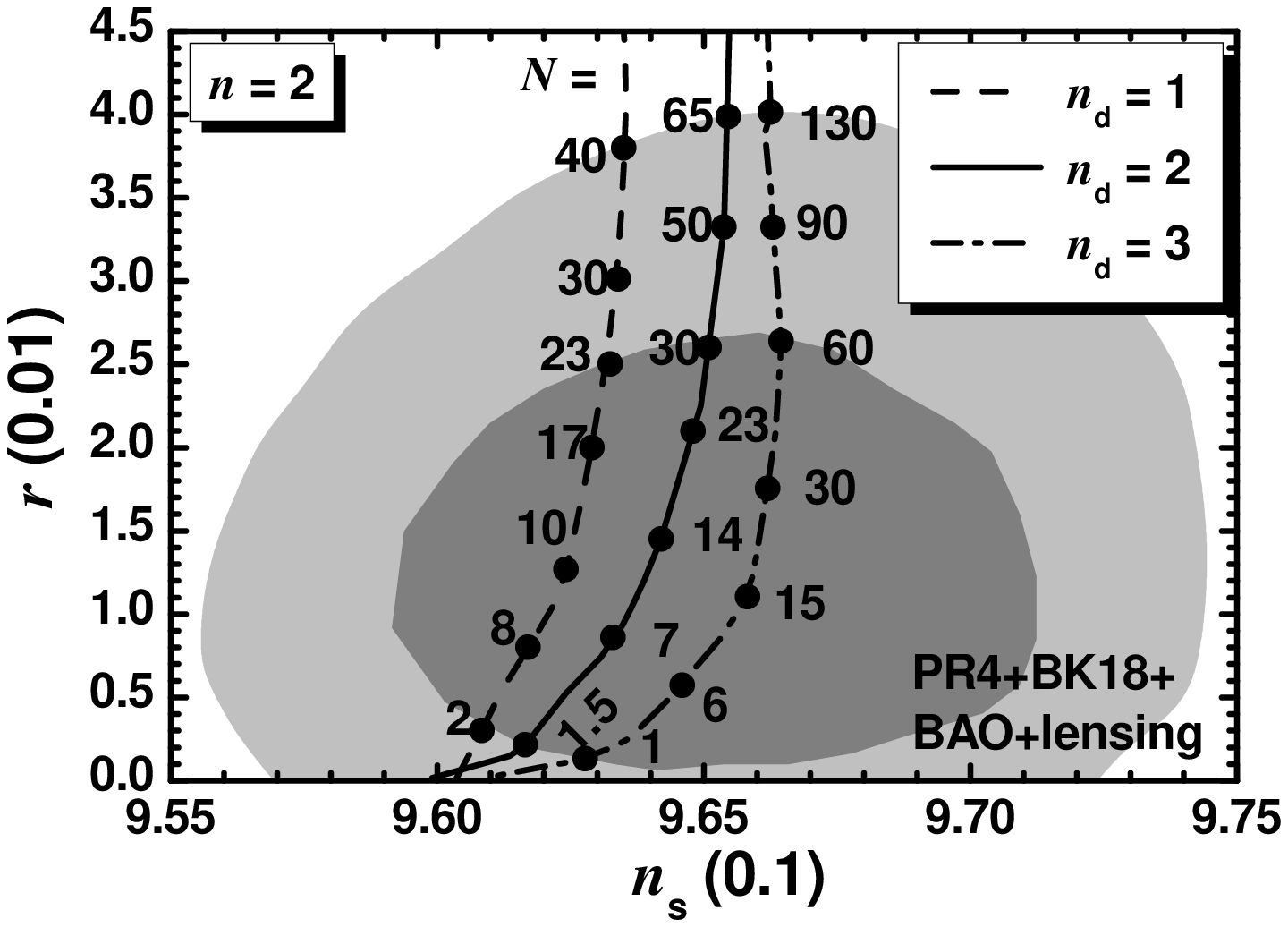,height=3.6in,angle=-90}
\hspace*{-1.2cm}
\epsfig{file=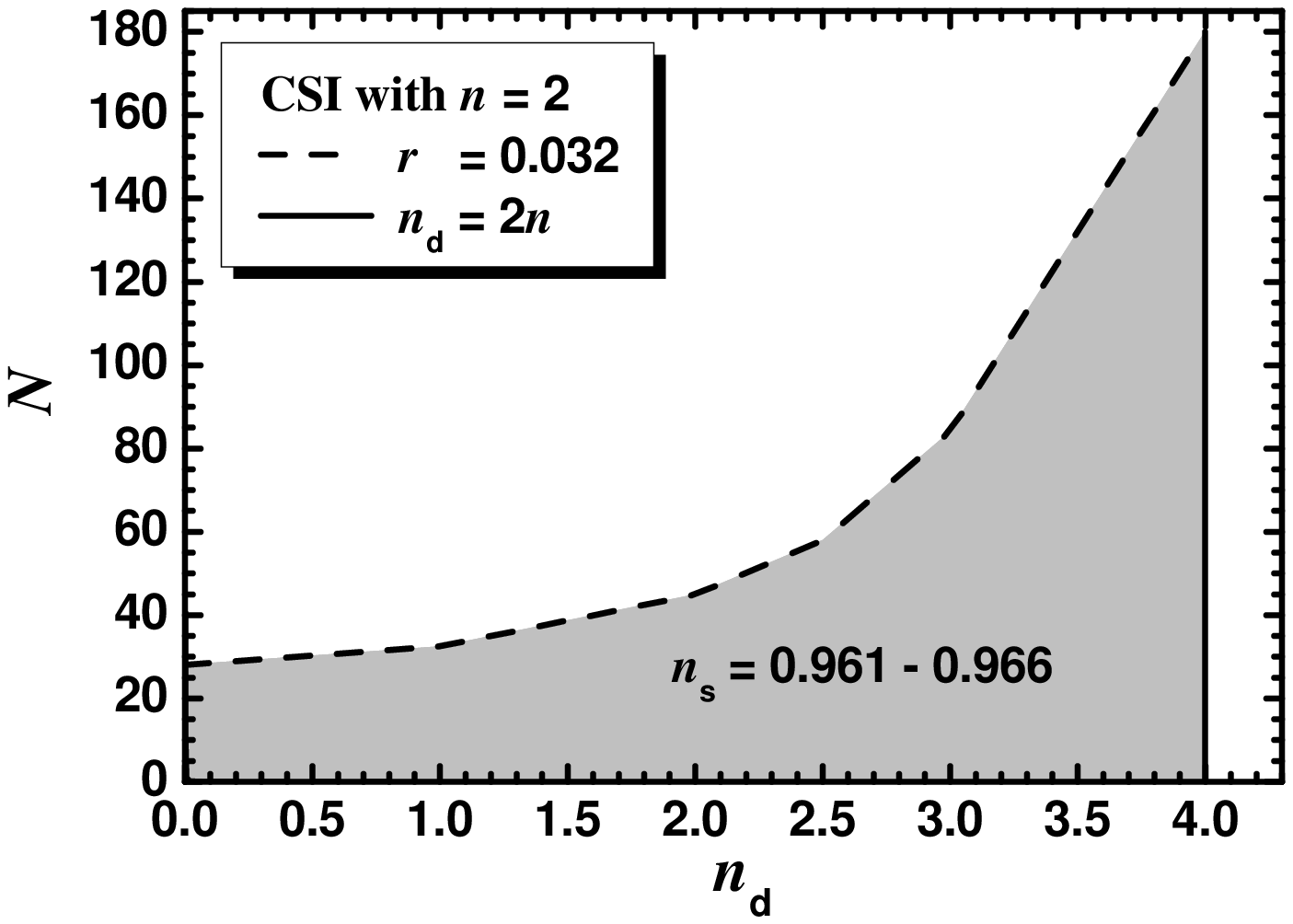,height=3.6in,angle=-90} \hfill
\end{minipage}\vspace*{-.3in}
\begin{flushleft}
\begin{tabular}[!h]{ll}
\hspace*{-.13in}
\begin{minipage}[t]{7.8cm}\caption[]{\sl\small Curves allowed by
\Eref{prob} in the $\ns-r$ plane for CSI with $n=2$, various $N$'s
indicated along them and $\nd=1$ (dashed line), $\nd=2$ (solid
line) or $\nd=3$ (dot-dashed line). The marginalized joint $68\%$
[$95\%$] c.l. regions \cite{gws} from PR4, {\sffamily\ftn BK18},
BAO and lensing data-sets are depicted by the dark [light] shaded
contours.}\label{fig1a}\end{minipage}
&\begin{minipage}[t]{7.5cm}\caption[]{\sl\small Allowed (shaded)
region as determined by Eqs.~(\ref{prob}) and (\ref{nspl}) in the
$\nd-\nm$ plane for CSI with $n=2$. The conventions adopted for
the boundary curves are also shown.}\label{fig1b}
\end{minipage}
\end{tabular}
\end{flushleft}\vspace*{-.11in}
\end{figure}

In this case we consider throughout $n=2$ which is motivated by
the quadratic potential which is usually encountered for
gauge-singlet superfields. The comparison of the model predictions
with data is displayed in \Fref{fig1a}, where we plot $r$ versus
$\ns$ for $\nd=1$ (dashed line), $\nd=2$ (solid line) or $\nd=3$
(dot-dashed line). The variation of $N$ is given along each curve.
We observe that the whole observationally favored range of $r$ is
covered varying $N$ whereas $\ns$ remains close to its central
value in \Eref{nspl}. As a consequence, an upper bound on $N$ can
be derived. This bound increases with $\nd$. Varying continuously
$N$ from $1$ until that maximal value, derived from the upper
bound on $r$ in \sEref{nspl}{b}, we obtain the shaded region in
\Fref{fig1b}. That upper bound, indicated by a dashed line, in
conjunction with the upper bound on $\nd$ inferred by \Eref{In}
and depicted by a solid line, delineate clearly the allowed
(shaded) parameter space of our model. In all, we find
\beqs\beq \label{res2}1\lesssim
N\lesssim180,~~~0\leq\nd\le3.99,~~0.961\lesssim\ns\lesssim0.966~~~\mbox{and}~~~1\lesssim\Dex/100\lesssim53\eeq
with $\wrh\simeq0$, $\as\simeq -(6.5-7.5)\cdot10^{-4}$ and
$\Ns\simeq(50-52)$. It is impressive that the $\Dex$ values are
much larger than the values derived in T-model Higgs inflation
analyzed in \cref{sor,jhepcs,polec} and therefore the present
model can be characterized as more natural. Note that the
naturalness of the model further requests $\nd\ll N$ since in this
regime the 't Hooft argument \cite{symm} suits better. Moreover,
$\ld$ and $\msn$ range as follows
\beq3.6\cdot10^{-5}\lesssim\ld\lesssim1.4\cdot10^{-4}
~~~\mbox{and}~~~1.6\lesssim\msn/10~\ZeV\lesssim4.7.\label{res2m}\eeq\eeqs
The maximal \msn\ values are obtained for the largest $\nd$ and
the minimal $N$ -- as deduced from the expression shown below
\Eref{msn}.

Representative values of model parameters, field values and
observables are given in the two leftmost columns of \Tref{tb2}
for $\nm=10$. This $\nm$ value gives $(\ns,r)$ in the current
``sweet'' spot of the dark shaded region in \Fref{fig1a}. We
notice the following: {\sf\small (i)} $\sgx$ and $\sgf$ are
subplanckian in accordance with the consideration of SUGRA as an
effective theory below $\mP=1$; {\sf\small (ii)} $\Dex$ increases
with $\nd$; {\sf\small (iii)} $\ld$ acquires a soft dependence
from $\nd$ not shown in its analytical expression in \Eref{lan};
{\sf\small (iv)}  $\Ldx<\mP$ is quite close to $\Hx$ and so the
effects of the renormalization-group running are negligible;
{\sf\small (v)} $w_{\rm rh}$ estimated by \Eref{wrh} is a little
lower than its naive value obtained by \Eref{wrha}; {\sf\small
(vi)} the values derived from the analytical expressions of
\Sref{hiobs} and written in italics are quite reliable for
$\nd=1$.

\begin{table}[!t]
\begin{center}
{\small
\begin{tabular}{|c||c|c|c|c|c|c|}\hline
{Model:}
&\multicolumn{2}{|c|}{CSI}&\multicolumn{2}{|c|}{HSI}&\multicolumn{2}{|c|}{HSI}\\\hline
$n$&\multicolumn{2}{|c|}{$2$}&\multicolumn{2}{|c|}{$4$}&\multicolumn{2}{|c|}{$8$}\\\cline{2-7}
$\nd$&$1$&$3$&$1$&$7$&$1$&$15$\\\hline\hline
$\sgx/0.1\mP$&$9.43~\{\it 9.4\}$&$8.8$&{$9.76~\{\it 9.7 \}$}&$9.1$&$9.7~\{\it 9.9\}$&{$9.4$}\\
$\Dex (\%)$&$5.7$&$11.5$&{$2.4$}&$9$&$1.5$&{$7$}\\
$\sgf/0.1\mP$&{$2.6$}~\{\it 1.9\}&$2.2$&{$4.7~\{\it
2.6\}$}&{$3.2$}&$6.9~\{\it 3.3\}$&{$4.4$}\\\hline
$\wrh$&$-0.065$&$-0.14$&{$0.29$}&$0.27$&$0.5$&{$0.49$}\\
$\Ns$&$51.5~\{\it 55\}$&$50.9$&{$56.5~\{\it 58
\}$}&$55.4$&$58.8~\{\it 60\}$&{$58.1$}\\\hline
$\ld/10^{-5}$&$2.9~\{\it 2.2\}$&$4.7$&{$2.8$}&$16.3$&$2.7$&{$23.2$}\\
$\Lex/10^{-5}\mP$&$2.3$&$2.8$&{$2.6$}&$1.6$&$2.5$&{$1.7$}\\
$\Hx/10^{-5}\mP$&$1.1$&$0.93$&{$1.1$}&$0.81$&$1.1$&{$0.72$}\\\hline
$\ns/0.1$&$9.62~\{\it 9.6\}$&$9.65$&{$9.64~\{\it
9.63\}$}&$9.67$&$9.66~\{\it
9.65\}$&{$9.68$}\\
$-\as/10^{-4}$&$7.1~\{\it 8.2\}$&$6.6$&{$6.3~\{\it 6.7\}$}&$5.8$
&$5.8~\{\it
6.2\}$&{$5.9$}\\
$r/10^{-2}$&$1.3~\{\it 1.4\}$&$0.8$&{$1.1~\{\it
1.2\}$}&$0.63$&$1.1~\{\it 1.1\}$&{$0.05$}\\\hline
$M$&\multicolumn{2}{|c|}{$-$}&\multicolumn{2}{|c|}{$6.4~\YeV$}&\multicolumn{2}{|c|}{$16.8~\ZeV$}\\\cline{2-7}
$\msn$&22.6~\ZeV&36.4~\ZeV&$79.7~\EeV$&{$46~\EeV$}&$1.1~\PeV$&{$89.6~\PeV$}\\\hline
\end{tabular}}
\end{center}
\caption{\sl\small Parameters, field values and observables
allowed by \eqs{prob}{nspl} for CSI with $n=2$ and HSI with
$\vev{\mbl}=\mgut$ and (i) $n=4$ or (ii) $n=8$. In all cases we
take $\nm=10$. Values in square brackets are obtained from our
analytical expressions in \Sref{hiobs}.} \label{tb2}\end{table}

\subsubsection{SI with a Higgs Field (HSI).}\label{res2b}

In this case we consider two representative values, $n=4$ and
$n=8$, which are appropriate for the self-consistency of $W$ in
\Eref{whi}. We also fix throughout $M$ imposing the GUT condition
in \Eref{Mg}. Since this is indistinguishable for both $K$'s
considered in \Sref{hi1}, our results are identical for both
cases. However, our results are valid for any other $M$ value
provided that $M\ll\mP$.

Several characteristic inputs and outputs for HSI with the
aforementioned $n$ values are listed in the central and the
rightmost columns of \Tref{tb2}. We take $N=10$ and $\nd=1$ and
$\nd=2n-1$ for both selected $n$ values. Recall that viable HSI
requires $\nd<2n$ -- see \eqs{Nhi}{In}. The remarks done in the
end of \Sref{res2a} regarding the findings of \Tref{tb2} are valid
for HSI too. Nonetheless, in this case we present also the $M$ and
$\msn$ values which are estimated via \eqs{Mg}{msn}
correspondingly. We see that both mass parameters decrease as $n$
increases and only $\msn$ develops a dependence on $\nd$ as
expected from the equations above.

One notable feature of our proposal is the fact that SI takes
place for subplanckian $\sg$ values. The naive assessment that
this achievement is not consistent with the chaotic character of
SI is not appropriate since $\sg$ does not coincide with the
canonically normalized inflaton, $\se$. If we take into account
the $\sg-\se$ relation of \Eref{tmd} we can easily infer that
$\se$ acquires transplackian values for $\sg<1$ and so SI is
rendered feasible -- cf. \cref{R2r, nIG, nIGpl, su11, su11a, ighi,
polec, sor}. To clarify further this key feature of our models, we
comparatively plot $\Vhi$ in \Eref{vhi} as a function of $\sg$ in
\sFref{fig4}{a} and $\se$ in \sFref{fig4}{b} for $N=10$, $n=4$ and
$\nd=1$ (black lines) and $\nd=7$ (gray lines).

From \sFref{fig4}{a} we see that $\Vhi$ for both $\nd$ values has
a parabolic-like slope for $\sg<1$. On the contrary, in
\sFref{fig4}{b} $\Vhi$ experiences a stretching for $\se>1$ which
results to the well-known plateau of SI for $\se\gg1$ -- see e.g.
\cref{su11a}. The observationally relevant inflationary period is
limited between the two $\sg$ values $\sgf$ and $\sgx$ which are
given in the two middle columns of \Tref{tb2} and are depicted in
\sFref{fig4}{a}. These values are enhanced, as advocated above,
and indicated in \sFref{fig4}{b}. Namely, solving \Eref{tmd} w.r.t
$\se$ we can estimate $\sef=2.3$ and $\sex=9.9$ for $\nd=1$ and
$\sef=1.5$ and $\sex=6.8$ for $\nd=7$. In the inset of
\sFref{fig4}{a} shown is also the structure of $\Vhi$ for low
$\sg$ values, responsible for the implementation of the $\bl$
phase transition -- see \Sref{hi3}.

\begin{figure}[!t]\vspace*{-.12in}
\hspace*{-.19in}
\begin{minipage}{8in}
\epsfig{file=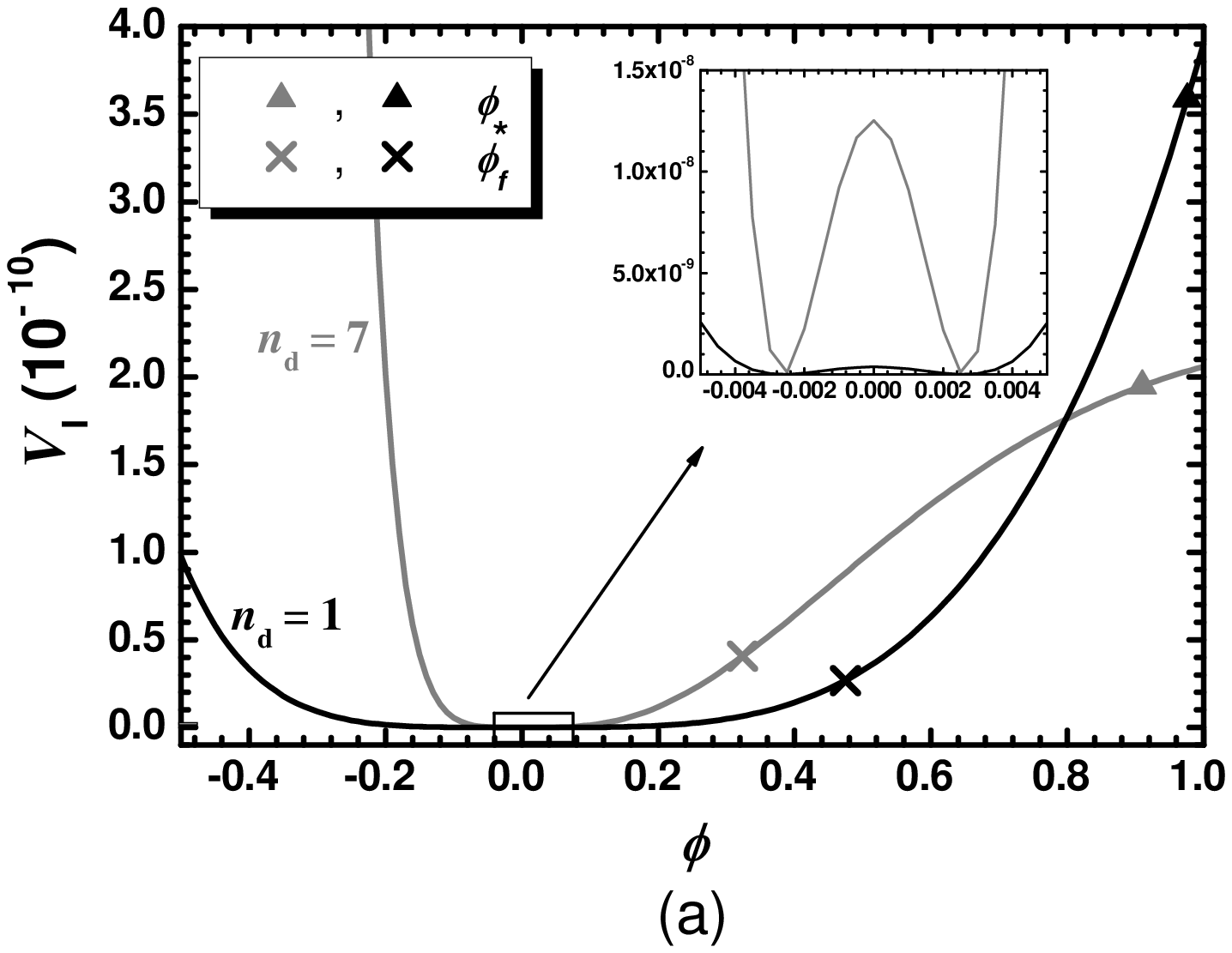,height=3.6in,angle=-90}
\hspace*{-1.2cm}
\epsfig{file=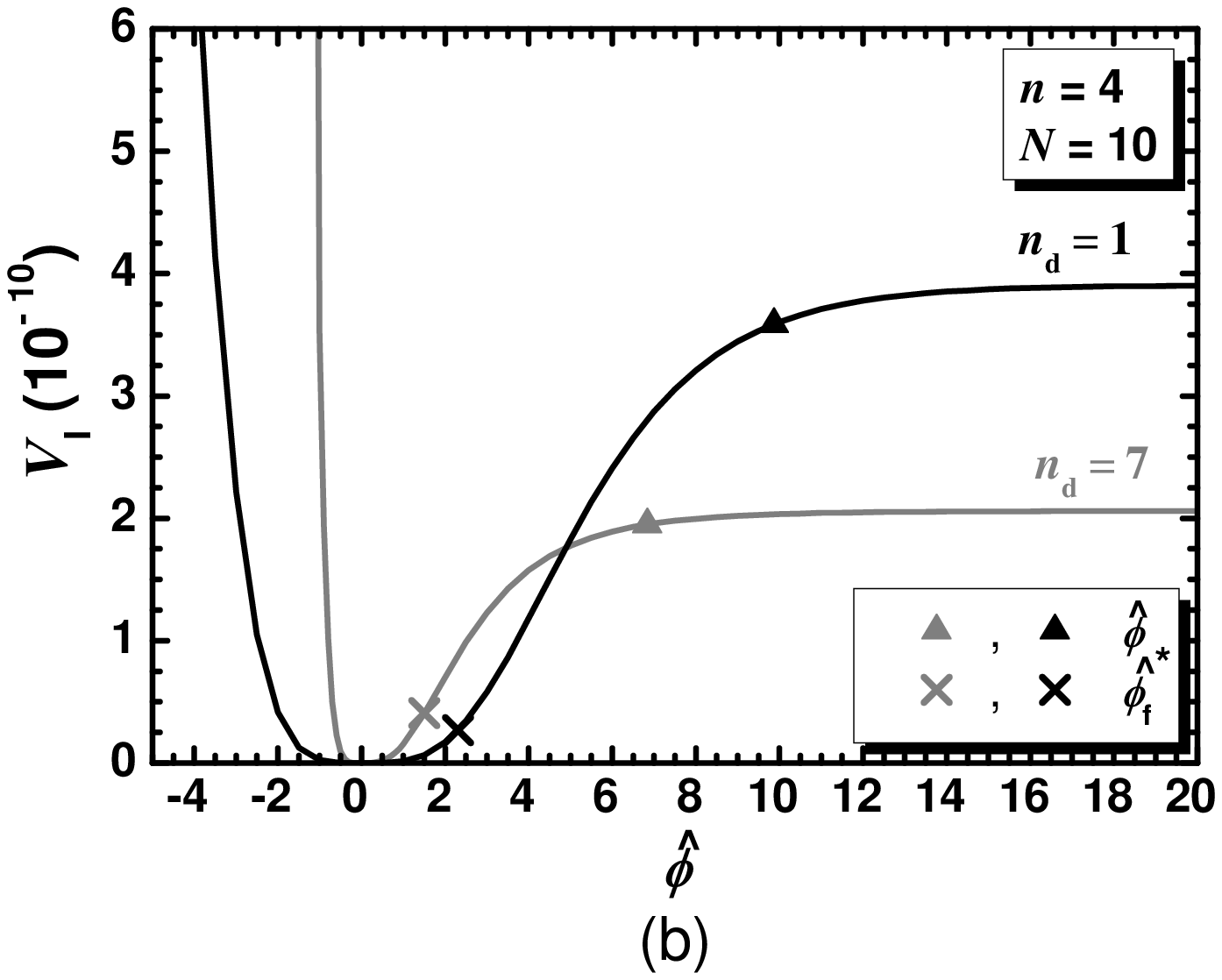,height=3.6in,angle=-90} \hfill
\end{minipage}
\caption[]{\sl\small Inflationary potential $\Vhi$ as a function
of {\sffamily\ftn (a)} $\sg$ and {\sffamily\ftn (b)} $\se$ fixing
$\vev{\mbl}=\mgut$. We consider HSI with $\nm=10$, $n=4$ and
$\nd=1$ (black lines) or $\nd=7$ (gray lines). Values
corresponding to $\sgx$ and $\sgf$ {\sffamily\ftn (a)} or $\se$
and $\sef$ {\sffamily\ftn (b)} are depicted. Shown is also the
low-$\sg$ behavior of $\Vhi$ in the inset {\sffamily\ftn (a)}.
}\label{fig4}
\end{figure}

From \sFref{fig4}{b} we remark that for both $\nd$ values the
magnitudes of the two plateaus are of the order $10^{-10}$ which
are similar to that obtained in pure SI
\cite{R2r,nIG,su11a,ighi,unh, unh1, blst, epole}. However, these
are one order of lower than that obtained in \crefs{nIGpl}, where
$r$ is a little more enhanced. Indeed, as verified from the values
listed in \Tref{tb2}, the level of the inflationary plateau
increases with $r$. Moreover, the inflationary scale $\Vci^{1/4}$
turns out to be well below $\mP$ and so the semi-classical
approximation, used in our analysis, is perfectly valid. Note that
here $\mP$ is undoubtedly the ultraviolet cut-off scale of the
theory thanks to the absence of large coefficients in the $K$'s.
Recall that such large coefficients are used in models of
induced-gravity \cite{R2r,nIG,su11,su11a,ighi,blst} or non-minimal
\cite{unh, unh1} inflation and the aforementioned scale has to be
determined after an expansion of ${\cal A}$ in \Eref{action1}
about $\vev{\sg}$.

In order to delineate the available parameter space of HSI for the
two selected $n$ values we plot its predictions in the $\ns-r$
plane against the observational data -- see \Fref{fig2}. To
accomplish it, we enforce the constraints in \Eref{prob} varying
$N$ for several $\nd$ values. Namely, for both $n$ values
considered, we fix $\nd=1$ (dashed lines), $\nd=n$ (solid lines)
or $\nd=2n-1$ (dot-dashed lines). Comparing the structure of these
plots with that of \Fref{fig1a} we see that this is pretty stable.
The $\ns$ values lie close to its central value in \Eref{nspl}
with a slight augmentation with $\nd$. The $r$ values increase
with $N$ whose the maximum increases with $\nd$. Considering that
maximum on $N$ for any allowed $\nd$ we show in \Fref{fig3} for
the two considered $n$ values the allowed (shaded) regions in the
$\nd-N$ plane. The findings are similar to that in \Fref{fig1b}
with an decrease of the maximal $N$'s and an increase of the
maximal $\nd$'s depicted by a dashed and a solid line
respectively. Obviously, the maximal of the maximal $N$ values are
obtained at the intersection of the dashed and the vertical solid
lines.

Summarizing our results for $n=4$ -- see \sFref{fig3}{a} -- we
arrive at the following allowed ranges
\beqs\beq \label{res4} 1\lesssim
N\lesssim165,~~~0\leq\nd\leq7.99,~~0.964\lesssim\ns\lesssim0.968~~~\mbox{and}~~~1\lesssim\Dex/100\lesssim41\eeq
with $\wrh\simeq0.3$, $\as\simeq -(5.8-6.5)\cdot10^{-4}$ and
$\Ns\simeq(54.-56)$. Moreover, $M$ and $\msn$ range as follows
\beq3.6~\YeV\lesssim M\lesssim43~\YeV
~~~\mbox{and}~~~47~\EeV\lesssim\msn\lesssim1.8~\ZeV.\label{res4m}\eeq\eeqs
On the other hand, for $n=8$ -- see \sFref{fig3}{b} -- we obtain
\beqs\beq \label{res8} 1\lesssim
N\lesssim152,~~~0\leq\nd\leq15.99,~~0.964\lesssim\ns\lesssim0.969~~~\mbox{and}~~~1\lesssim\Dex/100\lesssim30\eeq
with $\wrh\simeq0.5$, $\as\simeq -(5.3-6.2)\cdot10^{-4}$ and
$\Ns\simeq(57-59)$. As regards the mass parameters,
\beq4.8~\ZeV\lesssim M\lesssim0.17~\YeV
~~~\mbox{and}~~~0.16~\PeV\lesssim\msn\lesssim3.4~\EeV.\label{res8m}\eeq\eeqs
In both \eqs{res4m}{res8m} the maximal values are obtained for the
lowest $N$ and the largest $\nd$ values whereas the minimal values
are achieved at the largest $N$ and the lowest $\nd$ values.

\begin{figure}[!t]\vspace*{-.12in}
\hspace*{-.19in}
\begin{minipage}{8in}
\epsfig{file=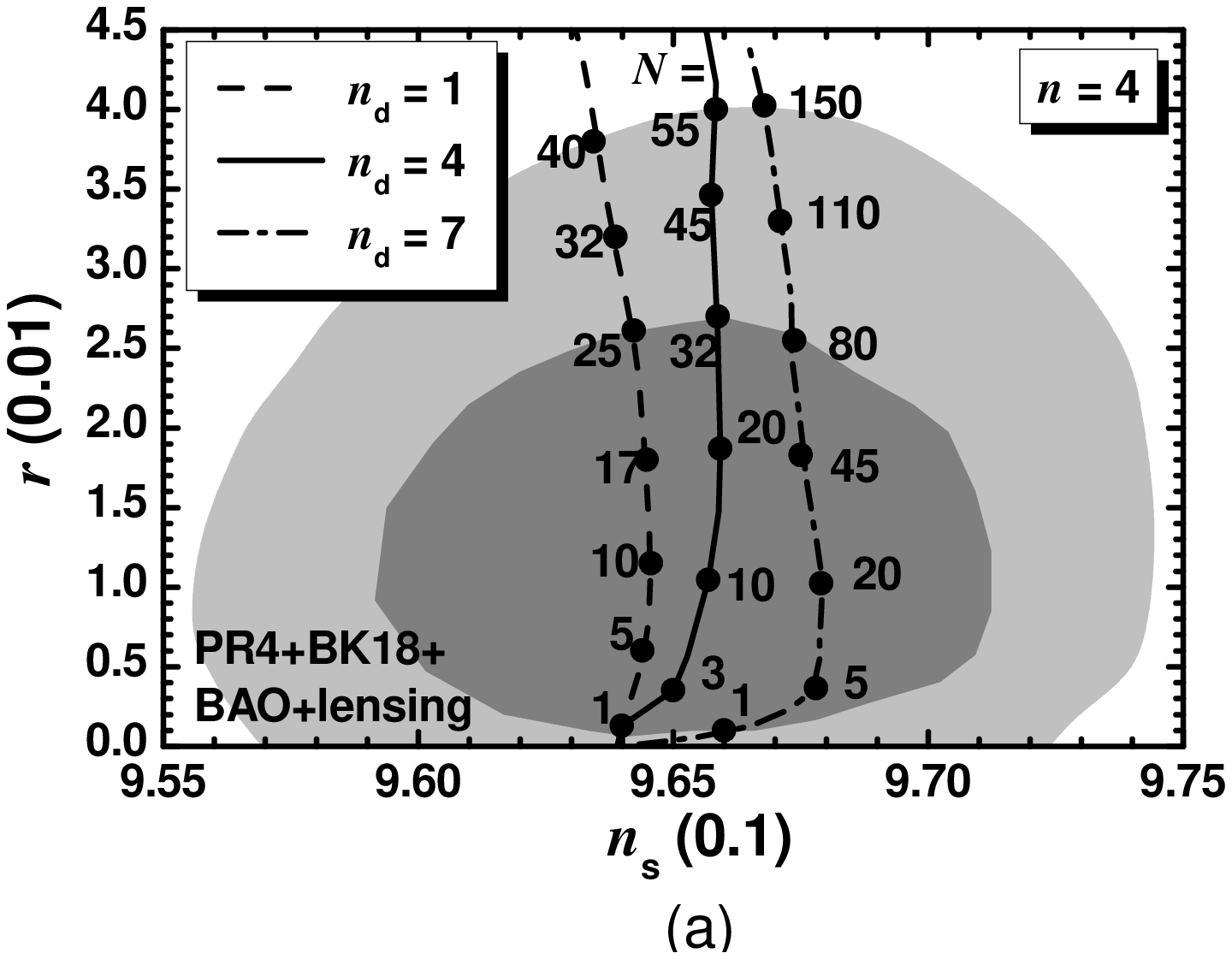,height=3.6in,angle=-90}
\hspace*{-1.2cm}
\epsfig{file=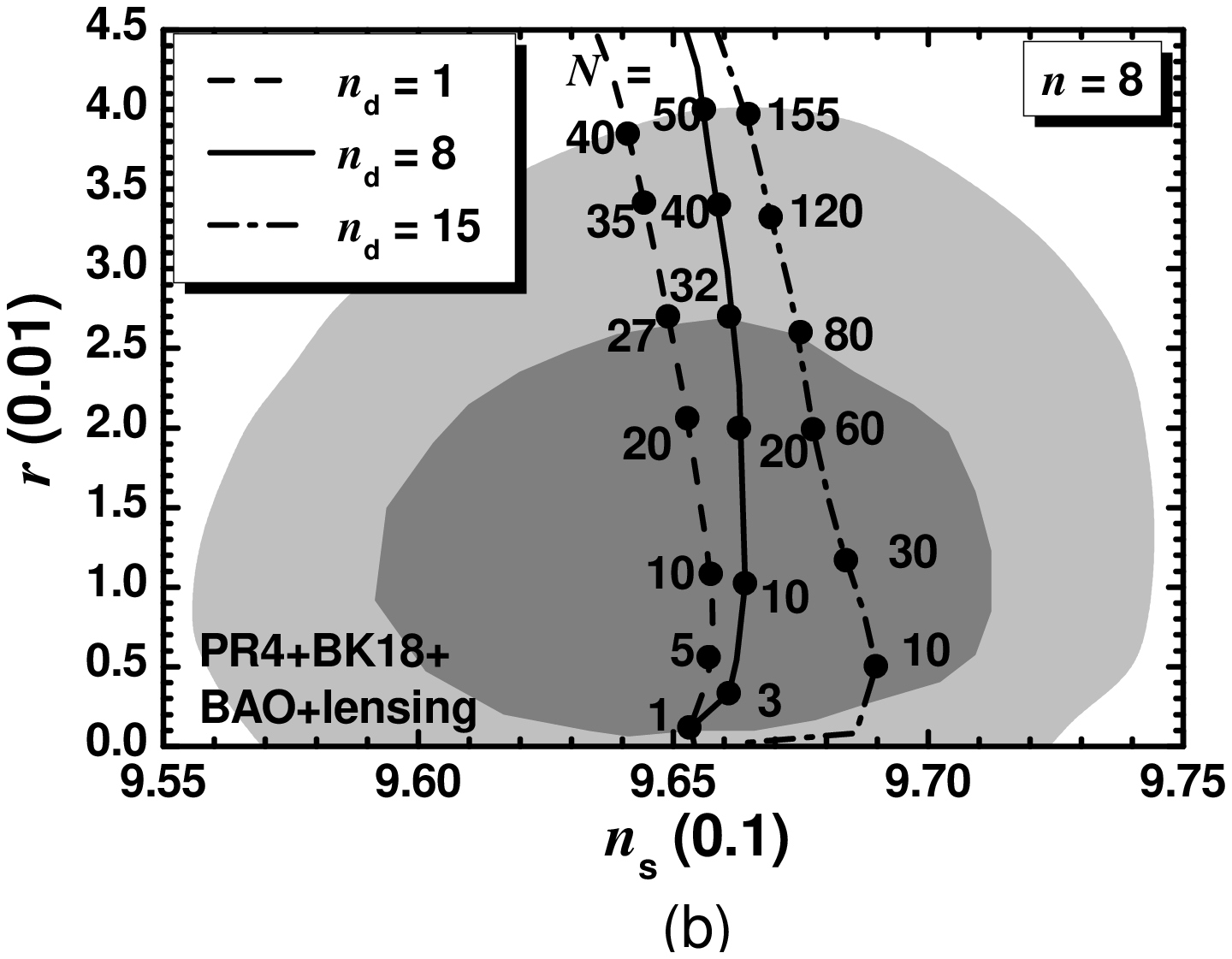,height=3.6in,angle=-90} \hfill
\end{minipage}
\hfill \caption{\sl\small  Curves allowed by \Eref{prob} in the
$\ns-r$ plane for HSI with $\vev{\mbl}=\mgut$ and various $\nm$'s
indicated along them. We take {\sffamily\ftn (a)} $n=4$ and
$\nd=1$ (dashed line), $\nd=4$ (solid line) or $\nd=7$ (dot-dashed
line) or {\sffamily\ftn (b)} $n=8$ and $\nd=1$ (dashed line),
$\nd=8$ (solid line) or $\nd=15$ (dot-dashed line). The shaded
regions are identified as in \Fref{fig1a}.}\label{fig2}
\end{figure}


\section{Conclusions}\label{con}

We presented a novel implementation of SI (i.e. Starobinsky
inflation) in the context of SUGRA confining ourselves to models
displaying a scalar potential shown in \Eref{vhi} and a kinetic
mixing in the inflaton sector with a pole of order two -- see
\Eref{tmd}. We considered two classes of models -- CSI, i.e.,
chaotic SI and HSI, i.e. Higgs SI -- depending on whether the
inflaton is included into a gauge-singlet or two gauge-non-singlet
fields. CSI and HSI are relied on the superpotentials in
\eqs{wci}{whi} respectively which respect an $R$ symmetry and
include an inflaton-accompanying field which facilitates the
establishment of SI. On the other hand, the \Kap's respect the $R$
and gauge symmetries and parameterize hyperbolic internal
geometries met in T-model inflationary settings. Namely, $K$ for
CSI is given in \Eref{tkbad} whereas for HSI we considered the two
distinct $K$'s shown in \Eref{khi}. The Higgsflaton in the last
case implements the breaking of a gauge $\bl$ symmetry at a scale
which may assume a value compatible with the MSSM unification. All
the models excellently match with the observations by restricting
the free parameters to reasonably ample regions of values. In
particular, $\ns$ lies close to its central observational value
while $r$ increases with $N$ -- see \Eref{res2} for CSI and
\eqs{res4}{res8} for HSI. The present data on $\rs$ and the
self-consistency of the models allows us to delineate the overall
allowed regions for selected values of the exponent $n$ in
\Eref{vhi} -- see \Fref{fig1b} and \ref{fig3}. The resulting
allowed margin of $\ns$ is extended from $0.961$ up to $0.969$,
depending on the chosen $n$, and therefore it is less restrictive
than the corresponding prediction of the traditional Starobinsky
model in \Eref{attr} -- i.e. $\ns\simeq0.963$. Hopefully, a more
accurate determination of $\ns$ and $r$ by future experiments
\cite{cmbs4,bird,prism,cmbpol} will assist us to single out the
most favorable one from the proposed models.

\begin{figure}[!t]\vspace*{-.12in}
\hspace*{-.19in}
\begin{minipage}{8in}
\epsfig{file=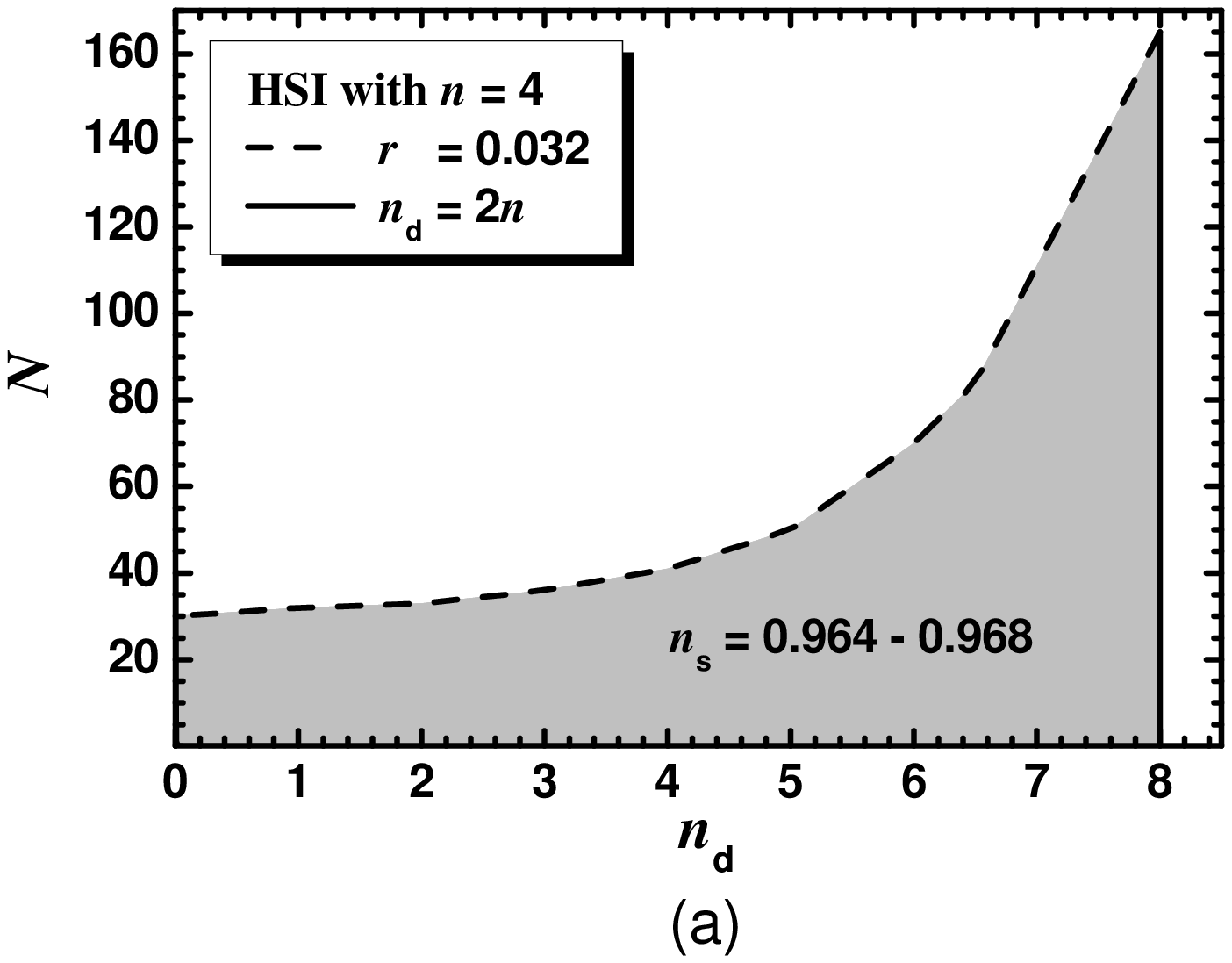,height=3.6in,angle=-90}
\hspace*{-1.2cm}
\epsfig{file=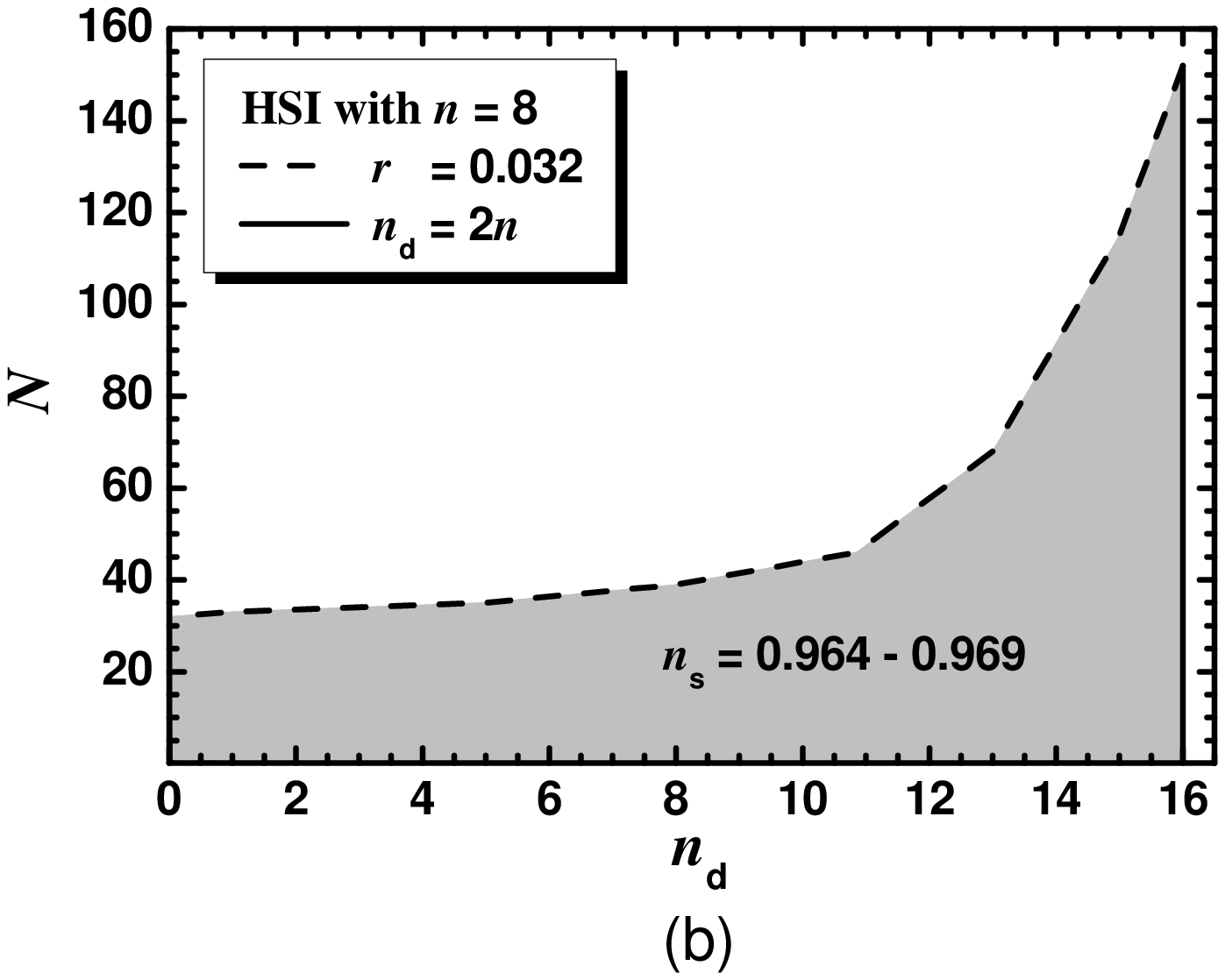,height=3.6in,angle=-90} \hfill
\end{minipage}
\hfill \caption{\sl\small  Allowed (shaded) region as determined
by Eqs.~(\ref{prob}) and (\ref{nspl}) in the $\nd-\nm$ plane for
HSI with $\vev{\mbl}=\mgut$ and {\sffamily\ftn (a)} $n=4$  or
{\sffamily\ftn (b)} $n=8$. The conventions adopted for the
boundary lines are also shown.}\label{fig3}
\end{figure}

The central message of our work is that SI is not exclusively
implemented by the E-model kinetic mixing in \Eref{emd}. It is
also attainable via T-model normalization in \Eref{tmd} if it is
considered in conjunction with the potential in \Eref{vhi}. The
method applied for the construction of our models can be extended
to other SUGRA models as those motivated by D branes
\cite{lindebr} or those which assist to obtain de Sitter vacua in
line with current LHC results on SUSY as in \cref{tkrefa,
unified}. Moreover, it can be employed for supersymmetrizing
models obtained adopting the Palatini approach to gravity
\cite{pal1, pal2, pal3}.

We expect that our models of HSI admit a post-inflationary
completion along the lines of \cref{R2r, blst, ighi, unh, sor2}
since the mass parameters in those models are similar to those
found in \eqs{res4m}{res8m}. As regards the ultraviolet
completion, it would be interesting to investigate if our models
belong to the string landscape or swampland \cite{conj1}. Note
that the swampland string conjectures are generically not
satisfied in SUGRA-based models \cite{conj2,conj3} but there are
suggestions \cite{conj4,conj5,conj6} which may work in our
framework too.




\paragraph*{\small \bf\scshape Acknowledgments} {\small  I
would like to thank I.~Ben-Dayan, S. Ketov and V. Zarikas for
useful discussions related to the Killing-adapted coordinates.}

\appendix{Shift Symmetry \& Hyperbolic \Ka\ Geometries}\label{app}

\renewenvironment{subequations}{%
\refstepcounter{equation}%
\setcounter{parentequation}{\value{equation}}%
  \setcounter{equation}{0}
  \def\theequation{A.\theparentequation{\sf\small \alph{equation}}}%
  \ignorespaces
}{%
  \setcounter{equation}{\value{parentequation}}%
  \ignorespacesafterend
}


\paragraph{\hspace*{.25cm}} We here demonstrate that the inflaton-sector \Kap s employed in our
work exhibit for $\nd=0$ a shift symmetry together with their
original hyperbolic structure, already extensively discussed in
\cref{tkref,tkrefa,sor}. To accomplish our goal we first find the
matrix form of the \Kme
\beq \mathbf{M}_K=\lf K_{\al\bbet} \rg ~~\mbox{with}~~z^\al=\bcs
\phc~~&\mbox{for}~~~K=\tka,\\ \phcb,
\phc~~&\mbox{for}~~~K=\tkaa~~\mbox{and}~~\tkbah. \ecs
\label{me}\eeq
We then introduce the so-called Killing-adapted coordinates
\cite{tkref,tkrefa} and express our $K$'s in terms of them
attempting to reveal a shift symmetry along the inflationary paths
of \Eref{inftr} or (\ref{inftrh}). We concentrate first on SI with
a gauge-singlet inflaton  -- see \Sref{app1} -- and then with a
gauge-non-singlet inflaton -- see \Sref{app2}.

\rhead[\fancyplain{}{ \bf \thepage}]{\fancyplain{}{\sc SI with
T-Model K\"ahler Geometries}} \lhead[\fancyplain{}{\sc
\hspace*{-0.1cm} Appendix A}]{\fancyplain{}{\bf \thepage}}
\cfoot{} 

\subsection{Shift Symmetry for CSI}\label{app1}

We concentrate on the following part of $\kbaad$ in \Eref{tkbad}
\beq\tka=-\nm\ln\frac{(1-|\phc|^2)}{(1-\phc^2)^{1/2}(1-\phc^{*2})^{1/2}},\label{tka}\eeq
which parameterizes $SU(1,1)/U(1)$ with curvature ${\sf
R}_{11}=-{2}/{N}$. The \Kme\ is a trivial $1\times1$ matrix with
element
\beq \mathbf{M}_{11}=N/(1-|\Phi|^2)^2. \label{ma}\eeq
We then introduce the superfield $\psc$ via the relation
\beq \phc=\tanh\frac{\psc}{\sqrt{2N}}.\label{Xdef}\eeq
Note that $\psc$ coincides with canonically normalized inflaton in
\Eref{kin}. Inserting it in \Eref{tka}, $\tka$ can be brought into
the form
\beq \tka=-N\ln\cosh\frac{\psc-\psc^*}{\sqrt{2N}},\label{tkaX}\eeq
if we take into account the identities of the hyperbolic functions
\beq \cosh(x-y)=\cosh x\cosh y(1-\tanh x\tanh y)
~~\mbox{and}~~\cosh x=(1-\tanh x)^{-1/2}. \label{hypid} \eeq
From the expression in \Eref{tkaX}, it is clear that $\tka$ is
invariant under the shift symmetry
\beq \psc \to\ \psc+c
~~\mbox{with}~~c\in\mathbb{R}.\label{shift}\eeq
Therefore, $\tka$ turns out to be independent from the canonically
normalized inflaton, $\se$ in \Eref{tmd} which can be identified
as the real part of $\psc$.

\subsection{Shift Symmetry for HSI}\label{app2}

We specify the emergence of a shift symmetry in the two $K$'s used
for HSI in \Eref{tkhi}.

\subsubsection{\Ka\ Manifold \mnfaa.}\label{app2a}

We concentrate on the inflationary contribution to $\tkbaad$ in
\Eref{tkhi} which reads
\beq\tkaa=-\frac{\nm}{2}\ln\frac{(1-2|\phc|^2)(1-2|\phcb|^2)}{(1-2\phcb\phc)(1-2\phcb^*\phc^*)}\label{tkaa}\eeq
and parameterizes $\mnfaa=(SU(1,1)/U(1))^2$ with curvature ${\sf
R}_{(11)^2}=-{4}/{(N/2)}$. The \Kme\ can be represented as a
diagonal $2\times2$ matrix
\beq \mathbf{M}_{(11)^2}=N\diag\lf
(1-|\phc|^2)^{-2},(1-|\phcb|^2)^{-2} \rg.  \label{maa}\eeq
working along the lines of the previous section, we introduce two
superfields $\psc$ and $\pscb$ via the relations
\beq
\phc=\frac1{\sqrt{2}}\tanh\frac{\psc}{\sqrt{2N}}~~~\mbox{and}~~~\phcb=\frac1{\sqrt{2}}\tanh\frac{\pscb}{\sqrt{2N}}.\label{Xdefa}\eeq
Upon substitution into \Eref{tkaa}, $\tkaa$ can be brought into
the form
\beq
\tkaa=-\frac{N}{2}\ln\frac{\cosh\frac{\psc-\psc^*}{\sqrt{2N}}\cosh\frac{\pscb-\pscb^*}{\sqrt{2N}}}
{\cosh\frac{\psc-\pscb}{\sqrt{2N}}\cosh\frac{\pscb^*-\pscb^*}{\sqrt{2N}}}.\label{tkaaX}\eeq
Taking into account that along the inflationary trough in
\Eref{inftrh} $\phc=\phcb$ and so $\psc=\pscb$, the expression
above reduces to the following
\beq
\left.\tkaa\right|_{\mbox{\ftn\Eref{inftrh}}}=-\frac{N}{2}\ln\cosh^2\frac{\psc-\psc^*}{\sqrt{2N}}.\label{Xshiftaaa}\eeq
Consequently, $\kaa$ is invariant under the shift symmetry of
\Eref{shift} and independent from $\Re\psc=\Re\pscb$ i.e. $\se$ --
see \eqs{tmd}{kzz}.

\subsubsection{\Ka\ Manifold \mnfbah.}\label{app2b}

Here we focus on the inflationary contribution to $\tkbbad$ in
\Eref{tkhi} which reads
\beq\tkbah=-\nm\ln\frac{1-|\phc|^2-|\phcb|^2}{(1-2\phcb\phc)^{1/2}(1-2\phcb^*\phc^*)^{1/2}},\label{kba}\eeq
which parameterizes $\mnfbah=SU(2,1)/U(1)$ with curvature ${\sf
R}_{21}=-{6}/{N}$. The \Kme\ is a non-diagonal $2\times2$ matrix
\beq \mathbf{M}_{21}=\frac{\nm}{\ffm^2}
\mtta{1-|\phcb|^2}{\phc^*\phcb}{\phc\phcb^*}{1-|\phc|^2},
\label{mba}\eeq
where $\ffm$ is given in \Eref{ffs}. The introduction of the
Killing-adapted coordinates can be now performed after
diagonalizing $\mathbf{M}_{21}$. This can be done via a similarity
transformation involving an hermitian matrix $U_{21}$ as follows:
\beq \label{diagMk} U_{21}^\dagger\mathbf{M}_{21} U_{21}
=\diag\lf\kpp,\kpm\rg\>\>\>\mbox{with}\>\>\>
U_{21}={1\over\sqrt{|\phc|^2+|\phcb|^2}}\mtt{|\phcb|\phc^*/\phcb^*}{-|\phc|\phcb/\phc}{|\phcb|}{|\phc|}.
\eeq
The eigenvectors and eigenvalues of $\mathbf{M}_{21}$ are given
respectively by
\beq \label{eig} \stl{\dot \phc_+}{\dot \phc_-} =U_{21}^\dagger
\stl{\dot \phc}{\dot \phcb}~~~\mbox{and}~~~\bcs \kpp=\nm/\ffm^2,
\\ \kpm=\nm/\ffm.\ecs \eeq
It is very difficult, if not impossible, to integrate the
relations above so as to determine generically \phc\ and \phcb\ in
terms of $\phc_\pm$. Therefore we are not able to obtain a generic
formula for $\tkbah$ as done in \Eref{tkaaX} for $\tkaa$. However,
confining ourselves along the direction in \Eref{inftrh} and
integrating the relevant relations in \Eref{eig} w.r.t the cosmic
time we find
\beq \label{phpm}\php=(\phcb+\phc)/{\sqrt{2}}~~~\mbox{and}~~~
\phm=(\phcb-\phc)/{\sqrt{2}}. \eeq
Solving the system above w.r.t $(\phcb, \phc)$ and taking into
account that $\vevi{\phm}=0$ -- see \eqs{hpara}{inftrh} -- we
obtain
\beq\left.\tkbah\right|_{\mbox{\ftn\Eref{inftrh}}}=-\nm\ln\frac{1-|\php|^2}{(1-\php^2)^{1/2}(1-\php^2)^{1/2}}.\label{vevkba}\eeq
We introduce again a new holomorphic variable $\psc_+$ via the
relation
\beq \php=\tanh\frac{\psc_+}{\sqrt{2N}}.\label{Xdefba}\eeq
During HSI $\psc_+$ becomes the real canonical variable $\se$ --
see \eqs{tmd}{kzz}. Inserting it in \Eref{vevkba}, it can be
brought into the form
\beq
\left.\tkbah\right|_{\mbox{\ftn\Eref{inftrh}}}=-N\ln\cosh\frac{\psc_+-\psc_+^*}{\sqrt{2N}}.\label{tkbaX}\eeq
This result manifests the invariance of $\tkbah$ under the
transformation of \Eref{shift} with $\psc$ replaced by $\psc_+$
and so $\tkbah$ is independent from $\Re\psc_+=\se$ -- see
\eqs{hpara}{phpm}.

\def\ijmp#1#2#3{{\sl Int. Jour. Mod. Phys.}
{\bf #1},~#3~(#2)}
\def\plb#1#2#3{{\sl Phys. Lett. B }{\bf #1}, #3 (#2)}
\def\prl#1#2#3{{\sl Phys. Rev. Lett.}
{\bf #1},~#3~(#2)}
\def\rmp#1#2#3{{Rev. Mod. Phys.}
{\bf #1},~#3~(#2)}
\def\prep#1#2#3{{\sl Phys. Rep. }{\bf #1}, #3 (#2)}
\def\prd#1#2#3{{\sl Phys. Rev. D }{\bf #1}, #3 (#2)}
\def\npb#1#2#3{{\sl Nucl. Phys. }{\bf B#1}, #3 (#2)}
\def\npps#1#2#3{{Nucl. Phys. B (Proc. Sup.)}
{\bf #1},~#3~(#2)}
\def\mpl#1#2#3{{Mod. Phys. Lett.}
{\bf #1},~#3~(#2)}
\def\jetp#1#2#3{{JETP Lett. }{\bf #1}, #3 (#2)}
\def\app#1#2#3{{Acta Phys. Polon.}
{\bf #1},~#3~(#2)}
\def\ptp#1#2#3{{Prog. Theor. Phys.}
{\bf #1},~#3~(#2)}
\def\n#1#2#3{{Nature }{\bf #1},~#3~(#2)}
\def\apj#1#2#3{{Astrophys. J.}
{\bf #1},~#3~(#2)}
\def\mnras#1#2#3{{MNRAS }{\bf #1},~#3~(#2)}
\def\grg#1#2#3{{Gen. Rel. Grav.}
{\bf #1},~#3~(#2)}
\def\s#1#2#3{{Science }{\bf #1},~#3~(#2)}
\def\ibid#1#2#3{{\it ibid. }{\bf #1},~#3~(#2)}
\def\cpc#1#2#3{{Comput. Phys. Commun.}
{\bf #1},~#3~(#2)}
\def\astp#1#2#3{{Astropart. Phys.}
{\bf #1},~#3~(#2)}
\def\epjc#1#2#3{{Eur. Phys. J. C}
{\bf #1},~#3~(#2)}
\def\jhep#1#2#3{{\sl J. High Energy Phys.}
{\bf #1}, #3 (#2)}
\def\prdn#1#2#3#4{{\sl Phys. Rev. D }{\bf #1}, no. #4, #3 (#2)}
\def\jcapn#1#2#3#4{{\sl J. Cosmol. Astropart.
Phys. }{\bf #1}, no. #4, #3 (#2)}
\def\epjcn#1#2#3#4{{\sl Eur. Phys. J. C }{\bf #1}, no. #4, #3 (#2)}


\end{document}